\documentclass[preprints,pdftex,accept,moreauthors]{Definitions/mdpi} 

\firstpage{1} 
\pubvolume{1}
\issuenum{1}
\articlenumber{0}
\pubyear{2025}
\copyrightyear{2024}
\datereceived{29 November 2024} 
\datepublished{ } 
\hreflink{https://doi.org/} % If needed use \linebreak
\Title{Estimating Large Global Significances with a New Monte Carlo Extrapolation Method}

\Author{Liangliang Chen $^{1,\dagger}$, Yufei Chen $^{1,\dagger}$, Gerry Bauer $^{1}$, Leonard G. 
Spiegel $^{2}$, \\ Zhen Hu $^{3,*}$\orcidA{} and Kai Yi $^{1,*}$\orcidB{}}

\address{%
$^{1}$ \quad Ministry of Education Key Laboratory of NSLSCS, Institute of Physics Frontiers and Interdisciplinary Sciences, School of Physics and Technology, Nanjing Normal University, Nanjing 210023, China; liangliang.chen@cern.ch (L.C.); yufei.chen@cern.ch (Y.C.); bauergp@yahoo.com (G.B.) %MDPI: we added emails, please confirm.
\\
$^{2}$ \quad Fermi National Accelerator Laboratory, Batavia, IL 60510, USA; lenny@fnal.gov\\
$^{3}$ \quad Department of Physics \& Center for High Energy Physics, Tsinghua University, Beijing 100084, China}

\corres{Correspondence: zhenhu@tsinghua.edu.cn (Z.H.);  kai.yi@njnu.edu.cn (K.Y.)}

\firstnote{These authors contributed equally to this work. }  % Current address should not be the same as any items in the Affiliation section.

\abstract{In particle physics, it is needed to evaluate the possibility that excesses of events in mass spectra are due to statistical fluctuations as quantified by the standards of local and global significances. 
Without prior knowledge of a particle's mass, it is especially critical to estimate its global significance. The usual approach is to count the number of times a significance limit is exceeded in a collection of simulated Monte Carlo~(MC) ``toy experiments.'' To demonstrate this conventional method for global significance, we performed simulation studies according to a recent Compact Muon Solenoid (CMS) result to show its effectiveness. 
However, this counting method is not practical for computing large global significances. To address this problem, we developed a new ``extrapolation'' method to evaluate the global significance. 
We compared the global significance estimated by our new method with that of the conventional approach, and verified its feasibility and effectiveness. This method is also applicable for cases where only small toy MC samples are available. In this approach, the significance is calculated based on \emph{p}-values, assuming symmetrical Gaussian distributions.}

\keyword{local significance; global significance; toy MC; extrapolation method; G-V method; log-likelihood ratio} 

\begin{document}

%%%%%%%%%%%%%%%%%%%%%%%%%%%%%%%%%%%%%%%%%%

% The order of the section titles is different for some journals. Please refer to the "Instructions for Authors” on the journal homepage.

\section{Introduction}	
There are a variety of discoveries in every area of physics, and~each of them needs to be examined by statistical standards to determine their significance above random fluctuations. In~particle physics, local significance and global significance are two general standards to evaluate claims of a new~particle. 

This article first illustrates the traditional method with MC simulated events to 
evaluate the significance. We use a new particle $X(7100)$ reported by the CMS Collaboration~\cite{CMS_2024,CPC2024,SB2024} as an example to illustrate the calculation of the global significance. We also describe in detail a method to estimate the chance of reaching a higher significance with additional data. This is useful for projecting how much additional data are needed to reach a global significance higher than five standard deviations~(5$\sigma$) in cases when the significance of a measured signal falls short of this benchmark, such as the case of the CMS $X(7100)$. 
Note that due to the size of the predefined search window, the~global significance can always be adjusted. 
This method for determining the significance, and~the expected significance, is general and can be applied, with~appropriate modification, to~most~cases. 

The  conventional method  mentioned above of calculating  global significance has limitations when the significance is very high.  
For example, the~$Y(4140)$ structure discovered by the CDF experiment~\cite{CDF2009,IJMPA2013,CDF2011} was confirmed in 2014 by the CMS experiment~\cite{CMS2014} to have a local significance of 7.6$\sigma$, corresponding to a probability (\emph{p}-value) of $1.5 \times 10^{-14}$, which means one would need an enormous collection of $1.5\times10^{14}$ Monte Carlo experiments to have, on~average, one toy instance which fluctuates above 7.6$\sigma$. It is very difficult to compute a global significance by the brute-force method of directly counting the number of simulated experiments which have a significance level around 7.6$\sigma$. %EE: check meaning retained

In order to overcome this issue, we describe a new extrapolation method that we have developed to evaluate the global significance when searching for new peaks by fitting the $\chi^2$ distribution of modest  MC samples, and~extrapolating to the large significances of interest. 
We use CDF's $Y(4140)$ as a test case to verify its feasibility and effectiveness. Two approaches, the~conventional toy MC method and the new extrapolation method, are used to compute the global significance of the $Y(4140)$ (which had a local significance of 5.3$\sigma$~\cite{CDF2009,CDF2011}) and compare the two results. Within~the margin of uncertainties, 0.1$\sigma$ in this case, we demonstrate the validity of our new method. This approach can be applied in a broad range of situations when searching for unknown~particles.

As an additional cross-check, another method proposed by Gross and Vitells (the G-V method)~\cite{Trial} can also be applied to estimate large global significances  with a small set of Monte Carlo simulations. This will allow us to obtain the global significance from the G-V method and compare it with our extrapolation method, again using the  CDF $Y(4140)$ example. 
We also apply the two  methods to compute the global significance, for~the first time, of~the $Y(4140)$ observed by CMS, which has a much higher local significance ($7.6\sigma$).	
Finally, to~further demonstrate the universality of the extrapolation method across different experiments, we apply it to the $\chi_b(3P)$ resonance observed by ATLAS~\cite{ATLAS} and compare the results with those obtained using the G-V~method.

\section{Significance~Measurements}
\unskip
\subsection{Local and Global~Significances}
When a small localized excess of events is seen in a mass spectrum,  the~question arises whether this is a new particle, or~merely a statistical fluctuation. The~probability that the data fluctuated specifically at that observed mass to fake a signal is referred to as the local significance of this ostensible~peak.

When studying a known particle, with~known  mass and width, the~local significance is the appropriate measure for the observation. However, in~the case of searching for an unknown particle, it is not known where it might appear in the mass spectrum, and one must consider the possibility of statistical fluctuations occurring {\it anywhere} in the search window---and not just where a particular fluctuation was observed. This is called the ``look-elsewhere-effect'' and by accounting for it, one obtains the global~significance. 

\subsection{Calculating the Local~Significance}
\label{callocal}
In actual experimental data, the~signal and background components are described by their respective PDFs (probability density functions). Mass spectra are commonly fit using the unbinned log-likelihood method~\cite{Byron}
as this is generally the optimum approach. To~evaluate the significance of a possible peak, one performs two fits. In~the first, one considers only the existence of the background component PDFs---this is the null-hypothesis fit. The~second fit is the background plus signal peak fit. $L0$ is defined as null fit's negative likelihood (NLL) and $L1$ is defined as the likelihood with the signal peak included. The~log-likelihood value of a fit is a parameter to show how good a fit is to the data, and~lower value means a better fit—typically, the signal-hypothesis fit has a lower value. The~difference in the NLL, $2 \times (L0-L1)$, is called the log-likelihood ratio, and~the local significance of the observed level of event excess is $\sqrt{2 \times (L0-L1)}$. This will be further explained in detail through an actual example in Section~\ref{exacmslocal}. 

\subsection{Using Toy MC to Calculate the Global~Significance}
The global significance can be calculated by simulating the background-only component of the observed experimental distribution, and~look for how often fluctuations as large, or~larger, than~those seen in the data occur anywhere in the search range.
To do this, a ``toy MC'' is used. 
In each ``toy'' or ``pseudo''  experiment,  a ``data'' set is simulated,  corresponding to the amount and shape of the background-only data distribution observed in the real experiment. This simulation is referred to as a ``toy'' as it only simulates the final distribution of interest in contrast to a full detector simulation of the data.
 A large number of toys are generated simulating the distribution of interest, and~they are used to conduct null-hypothesis and signal-hypothesis fits to obtain the $L0$ and $L1$ values. For~any given toy, a local significance can be calculated for a limited region, as  explained in Section~\ref{callocal}. 

However, the~local significance estimation accounts for the probability of a fluctuation occurring at a predetermined mass value, and~the problem with this significance is that a fluctuation may appear anywhere in the search window.
In other words, the~location of the signal component function in the signal-hypothesis fit must freely float over the mass and width search range that was defined in advance. 
For each toy, the null and signal fits are conducted, and~$2 \times (L0-L1)_{MC}$ is computed, as well as $2 \times (L0-L1)_{real-data}$ for the log-likelihood ratio in the real experiment.
Then, by counting the number of cases where $2 \times |L0-L1|_{MC} > 2 \times |L0-L1|_{real-data}$,  the~probability (\emph{p}-value) of toy experiments having a likelihood ratio above that observed in the real data is computed. 
A symmetrical Gaussian distribution is used to convert this probability into a Gaussian equivalent for the number of standard deviations, i.e.,~for some standard deviation $X$,
the integral  of a normalized Gaussian function from $X$ to positive infinity is  equal to the observed \emph{p}-value---this $X$ is the corresponding global significance in standard deviations. The~details will be further explained in the example in Section~\ref{exacmsglobal}.

\section{Conventional Global Significance~Estimation}
\begin{figure}[H]
\centering
 {\includegraphics[width=10cm]{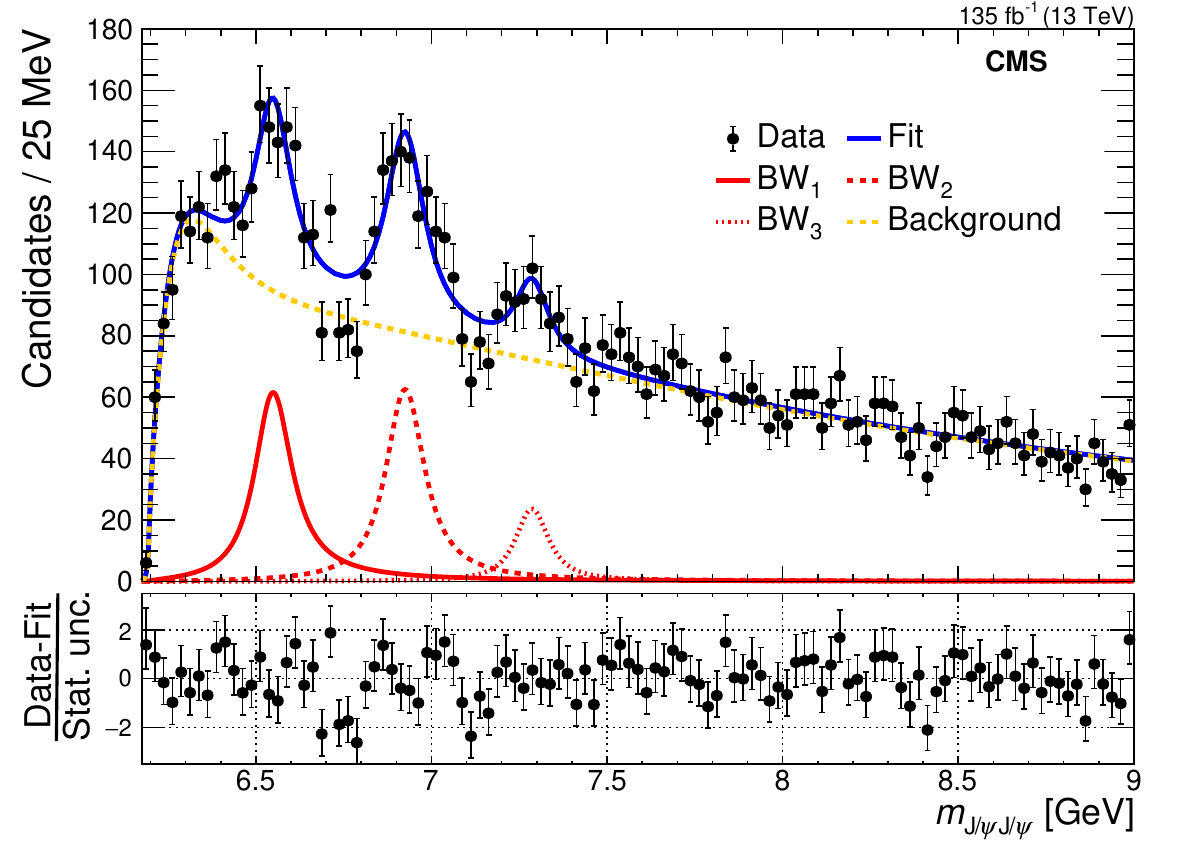}}
\caption{CMS  experiment's  $J/\psi J/\psi$ invariant mass spectrum fit with the no-interference model showing three structures~\cite{CMS_2024}.\label{cms21003}}
\end{figure}

Here, we use an actual experimental example to illustrate the conventional method for significance estimation. Recently, the~CMS Collaboration tentatively identified three exotic hadron candidates, referred to as 
$X(6600)$ ($m=6638^{+43+16}_{-38-31}$ MeV, $\Gamma=440^{+230+110}_{-200-240}$ MeV), 
$X(6900)$ ($m=6847^{+44+48}_{-28-20}$ MeV, $\Gamma=191^{+66+25}_{-49-17}$ MeV)
and $X(7100)$ ($m=7134^{+48+41}_{-25-15}$ MeV, $\Gamma=97^{+40+29}_{-29-26}$ MeV), 
in the $J/\psi J/\psi$ mass spectrum in proton--proton collisions at $\sqrt{s} = 13$ TeV~\cite{CMS_2024}, as~shown in Figure~\ref{cms21003} with their non-interference fit model.
They also used an interference model to fit their spectrum, but~we use the simpler non-interference case as an example for our test. The~local significances of $X(6600)$ and $X(6900)$ were far above 5 standard deviations. The~large significances leave no doubt that these are not statistical fluctuations. However, the~local significance of the $X(7100)$ was only approximately $4.1 \sigma$. Since the mass and width of $X(7100)$ were not predicted before it was found, the~evaluated local significance did not consider its unknown mass and width. In~this section, the~global significance of this potential new particle $X(7100)$ with unknown mass and width will be evaluated using simulated events, as~a demonstration of the conventional method to calculate global significance, as~described in Ref.~\cite{Kelly}.

\unskip

\subsection{Example of CMS Local~Significance}
\label{exacmslocal}
In this CMS example, the~background components are modeled with two threshold functions used in the CMS analysis~\cite{CMS:2022yhl}, while the signal component is described by a standard relativistic Breit--Wigner formula~\cite{CMS:2022yhl}:
\begin{eqnarray}
\begin{array}{rcl}
  BW(m;m_0,\Gamma_0) &=& 
  \displaystyle{\frac{\sqrt{m\Gamma(m)}}{m_0^2 - m^2 - i m\Gamma(m)}}, \\[8pt]
  \Gamma(m) &=& \displaystyle{\Gamma_0} \left(\frac{q}{q_0}\right)^{2L + 1}\frac{m_0}{m}\left(B^{\prime}_L(q, q_0, d)\right)^2 , \\[8pt]
  B^\prime_L(q, q_0, d) 
  &=& \displaystyle{\frac{q^{-L}B_L(q,d)}{q_0^{-L}B_L(q_0,d)}}
  = \left(\frac{q_0}{q}\right)^L\frac{B_L(q,d)}{B_L(q_0,d)}, \\[8pt]
  B_0(q, d) &=& 1, \\[8pt]
  z &=& (|q|d)^2, z_0 = (|q_0|d)^2, 
\end{array}
\label{eqn3}
\end{eqnarray}
where $q$ is the magnitude of momentum of a daughter in the resonance rest frame, the~parameter $L$ is the internal orbital angular momentum, and~CMS chose the simplest case of $L=0$ (S-wave) for their baseline fit. The~parameter $d$ is set to be 3 GeV$^{-1}$.

\begin{figure}[H]
\centering
 {\includegraphics[width=12cm]{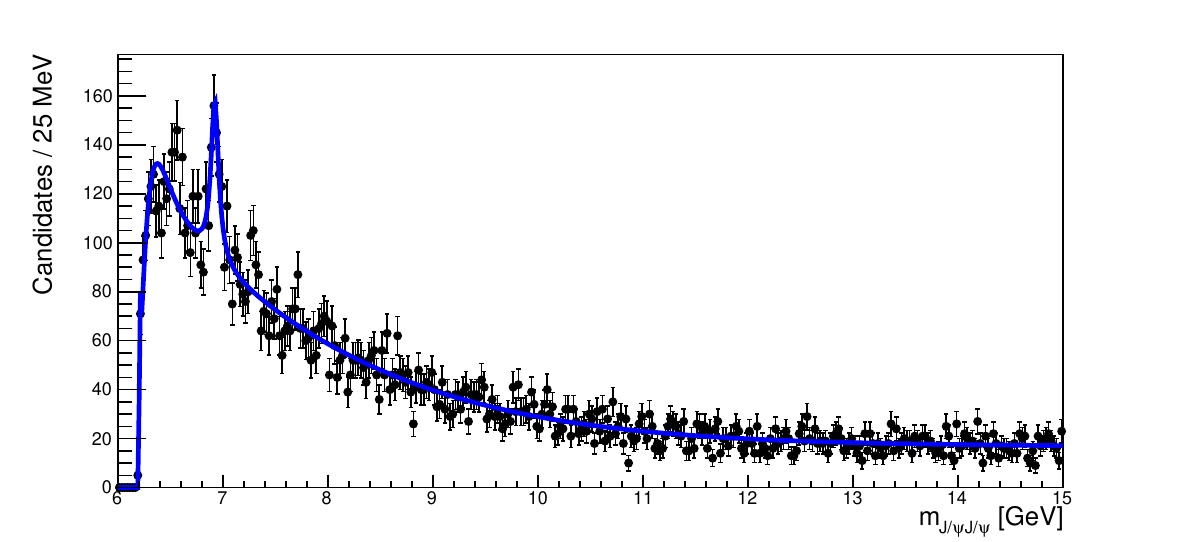}}
 {\includegraphics[width=12cm]{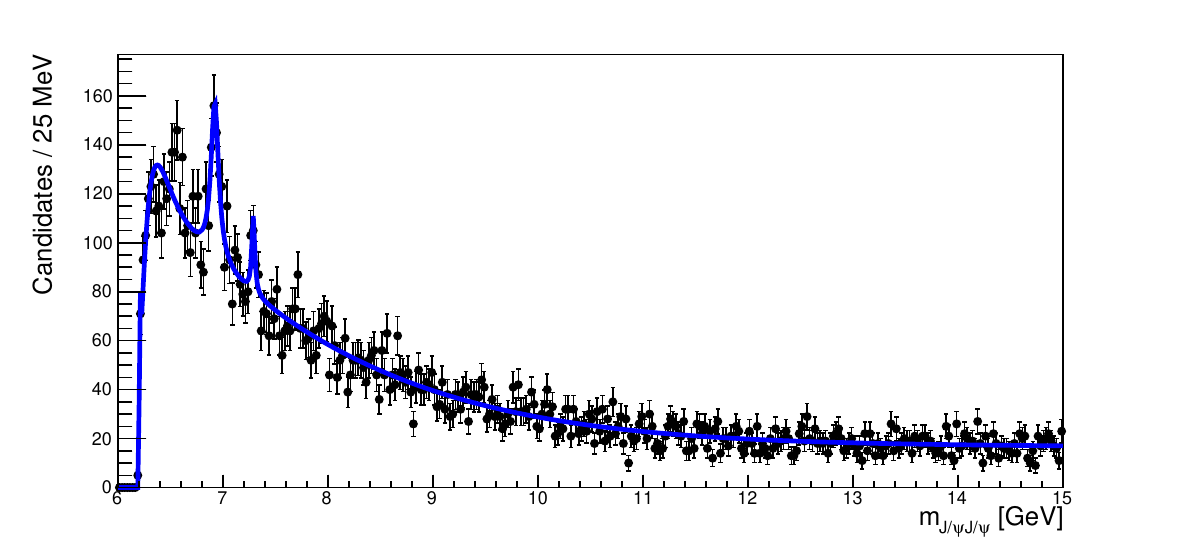}}
\caption{An example of a toy experiment simulating CMS's $J/\psi J/\psi$ mass spectrum along with a test fits. 
The upper panel shows the null-hypothesis fit to this toy, while the lower panel shows the signal-hypothesis fit the same toy. A~local significance of 4.4 standard deviations for the $X(7100)$ peak is obtained by comparing the likelihoods of the two fits. \label{toy7100}}
\end{figure}

The signal and background formulas are implemented using the ROOT/RooFit~\cite{root} \linebreak software [version 6.32.08] package as PDFs.
 Each PDF is used to generate simulated events according to its distribution.  
Each MC toy sample for the CMS mass spectrum is generated  with the following components, which are combined together for our test: 13000 background events, 500 $X(6600)$ events, 500 $X(6900)$ events, and~150 $X(7100)$ events. Here, the number of events for each component is approximately based on the CMS analysis~\cite{hepdata}. An~example of the mass spectrum generated for  a toy experiment is shown in Figure~\ref{toy7100}. 

To obtain the local significance of the $X(7100)$ peak for this toy,  a~log-likelihood fit to this simulated data is performed where the $X(7100)$ PDF is excluded, but~all other PDF components (including $X(6600)$ and $X(6900)$) are included in the fitting model. This fit is shown in the upper panel of Figure~\ref{toy7100}. The~$X(7100)$ mass and width are fixed to the values from the CMS result---this glosses over the fact that its mass and width were not known prior to CMS's analysis.
This fit is the null-hypothesis fit, and~$L0$ is the value of the minimized log-likelihood.
A signal fit is also conducted which has all components, including the $X(7100)$ PDF, and~which is shown in the lower panel of Figure~\ref{toy7100}, and~results in the likelihood value  $L1$.  The~difference between $L1$ and $L0$ determines the statistical size of a potential $X(7100)$ signal, which is quantified as a local significance by $\sqrt{2 \times (L0-L1)}$. The~significance value for the example shown in Figure~\ref{toy7100} is 4.4$\sigma$, which is close to the CMS result of 4.1$\sigma$.

\vspace{-6pt}
 
\subsection{Example of CMS Global~Significance}
\label{exacmsglobal}
\begin{figure}[H]
\centering
{\includegraphics[width=7.5cm]{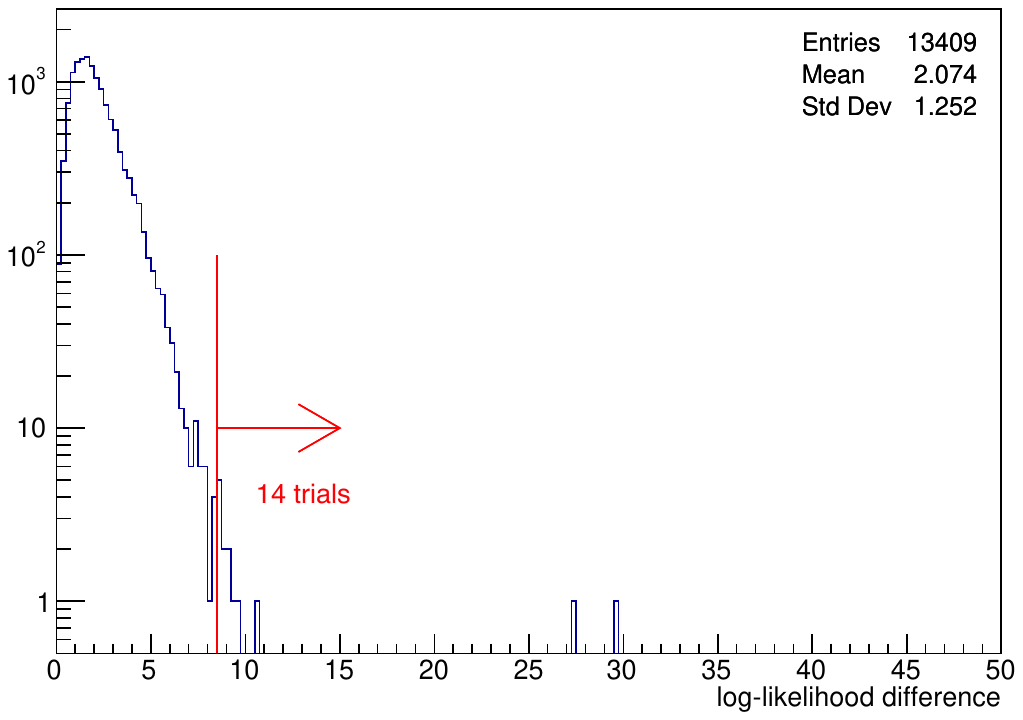}}
\caption{The log-likelihood difference, $(L0-L1)_{trial}$, of~13409 trials simulating the CMS case study. The~red line and arrow indicate  that there are 14 trials with a value greater than~8.5. \label{f2}}
\end{figure}

To calculate the global significance for the CMS $J/\psi J/\psi$ example, we performed a large number of pseudo-experiments. In each experiment, a~simulated background-only toy MC sample similar to the one in Figure~\ref{toy7100} but without $X(7100)$ was produced. As~suggested in the CMS analysis~\cite{CMS_2024}, we searched for fluctuations mimicking signals in the mass region between 7.05 and 7.8 GeV with a step size of 0.1 GeV, and with the width between 45 and 135 MeV---a range which spans one standard deviation of the value obtained in the CMS analysis---and scanned the width with a step size of 25 MeV.
The log-likelihood difference
$(L0-L1)_{trial}$ is computed for the largest fluctuation found in each case. 
Figure~\ref{f2} shows the distribution of $(L0-L1)_{trial}$ for these trials. In total, 14 out of the 13409 trials have a log-likelihood difference of over 8.5 (the value obtained in CMS data), so the \emph{p}-value is calculated as $14/13409=0.00104$.

\begin{figure}[H]
\centering
{\includegraphics[width=8.5cm]{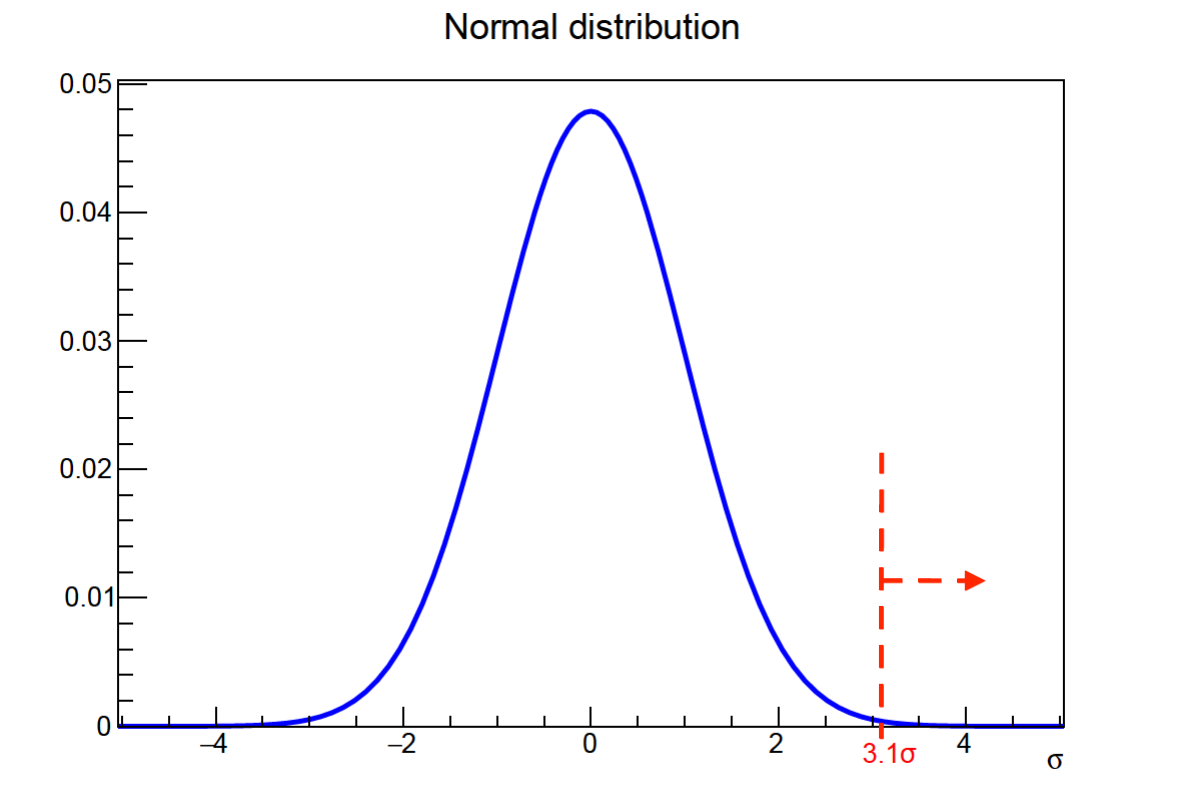}}
\caption{A normal distribution showing the $3.1 \sigma$ significance of the CMS example corresponding to the \emph{p}-value of 0.00104. The red arrow indicates the area with an integral value of 0.00104. 
\label{f3}}
\end{figure}

This small \emph{p}-value signifies a low probability of the excess  resulting from a random fluctuation. The~probability, 0.00104,  is converted  into a significance assuming the probability follows a symmetrical Gaussian distribution: the \emph{p}-value of 0.00104 corresponds to $3.1 \sigma$, i.e.,~0.00104 is the area of the integral of a normalized Gaussian distribution from $3.1 \sigma$ to positive infinity, as~illustrated in Figure~\ref{f3}. 

\vspace{-9pt}

\subsection{Expected Significance  with Additional~Data}
To verify if a newly found signal is real, it may be necessary to increase the data sample enough to surpass the desired statistical threshold.
The following question arises before conducting such an update:
 how much additional data are needed to combine with the original sample to reach 5$\sigma$ significance? For instance, using the above CMS example of a new signal, $X(7100)$,  with~a local significance of 4.1$\sigma$, a~rough estimate often used to estimate the expected significance of an updated analysis is that
 the  expected significance would be $4.1\sigma \times \sqrt{F}$, where $F$ is the fraction of the combined data size compared to the original data size. Thus, the expected significance reaches 5$\sigma$ when $F$ is 1.5 in this example; that is to say, the~4.1$\sigma$ new signal can reach 5$\sigma$ if the original data are combined with $50\%$ additional data.  This is a rough estimate, while in reality, the actual significance depends on many factors such as background shapes, fractional increase in background, etc.

\begin{figure}[H]
\centering
{\includegraphics[width=8cm]{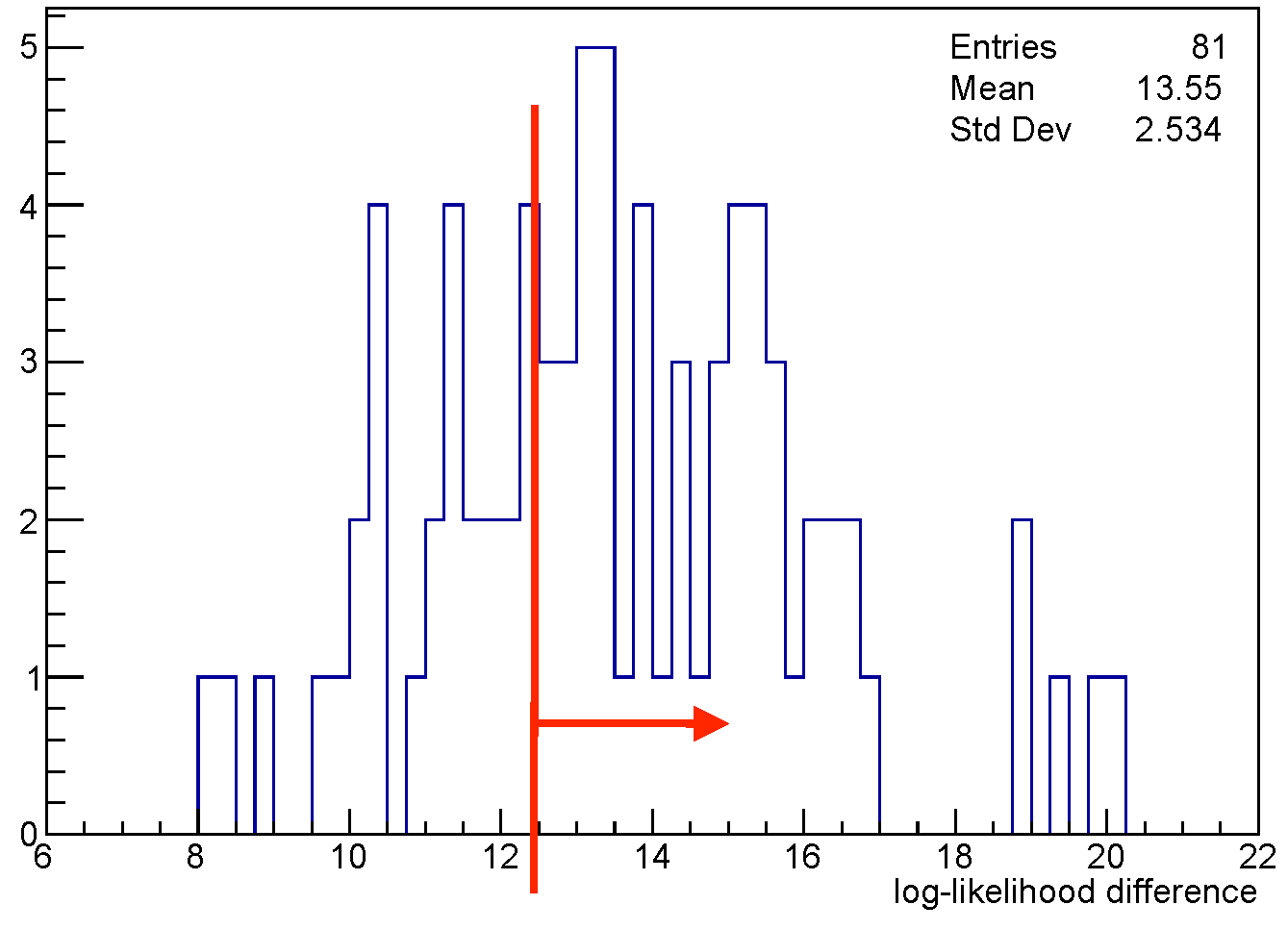}}
\caption{The log-likelihood difference distribution for a semi-simulated exercise to project the significance of CMS's observed $X(7100)$ signal. The~original sample is supplemented by a simulation of
50$\%$ more signal and 50$\%$ additional background. \label{f4}}
\end{figure}

Here, we illustrate a more complete approach to estimate the average expectation of reaching 5$\sigma$ given an existing 4.1$\sigma$ signal when an enlarged  data sample contains $50\%$ additional signal events and 25$\%$, 50$\%$, or~75$\%$ additional background events. 
In total,  50$\%$ of additional signal events %EE: check meaning retained
are generated according the signal PDF function, and~25$\%$, 50$\%$, and~75$\%$ of background events according to the background PDF function.
These simulated events are then combined with the original data and used to evaluate the local significance of the combined signal using the likelihood method described earlier. 
Figure~\ref{f4} shows the likelihood differences for 81 examples where
50$\%$ additional signal and 50$\%$ additional background are generated. 
In this distribution, 
 54 out 81 cases (66$\%$)  are at or above 5$\sigma$. Alternatively, the~probability is about 39$\%$ and 92$\%$ if the additional background combined is 75$\%$ and 25$\%$, respectively. 

This exercise illustrates that certain factors can alter the chance of reaching 5$\sigma$ when adding additional data, and~these effects can be estimated by using simulated events. 
To effectively design and optimize an analysis, it is crucial to estimate the amount of new data required to achieve statistical targets for updated analyses. Failing to determine this beforehand may introduce bias into the~results.

\section{Methodology: A New Extrapolation Method for Global~Significance}
\unskip
\subsection{Limitations of the Conventional Toy MC~Method}
\label{limitmc}
The conventional method of calculating global significance by directly counting the number of cases out of a large set of toy experiments, as~shown in the previous section, has serious~limitations. 

\begin{figure}[H]
\begin{center}
 \includegraphics[width=0.4\textwidth]{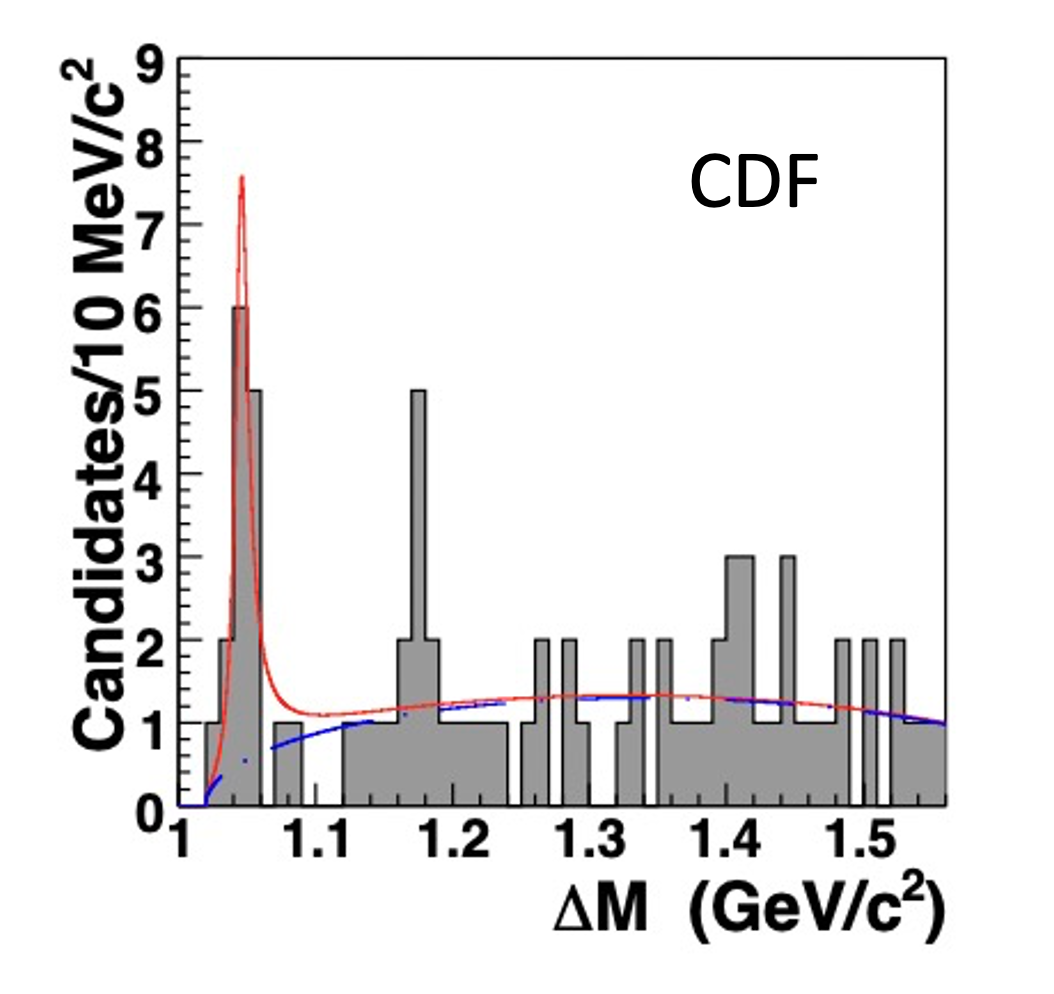}  
\includegraphics[width=0.5\textwidth]{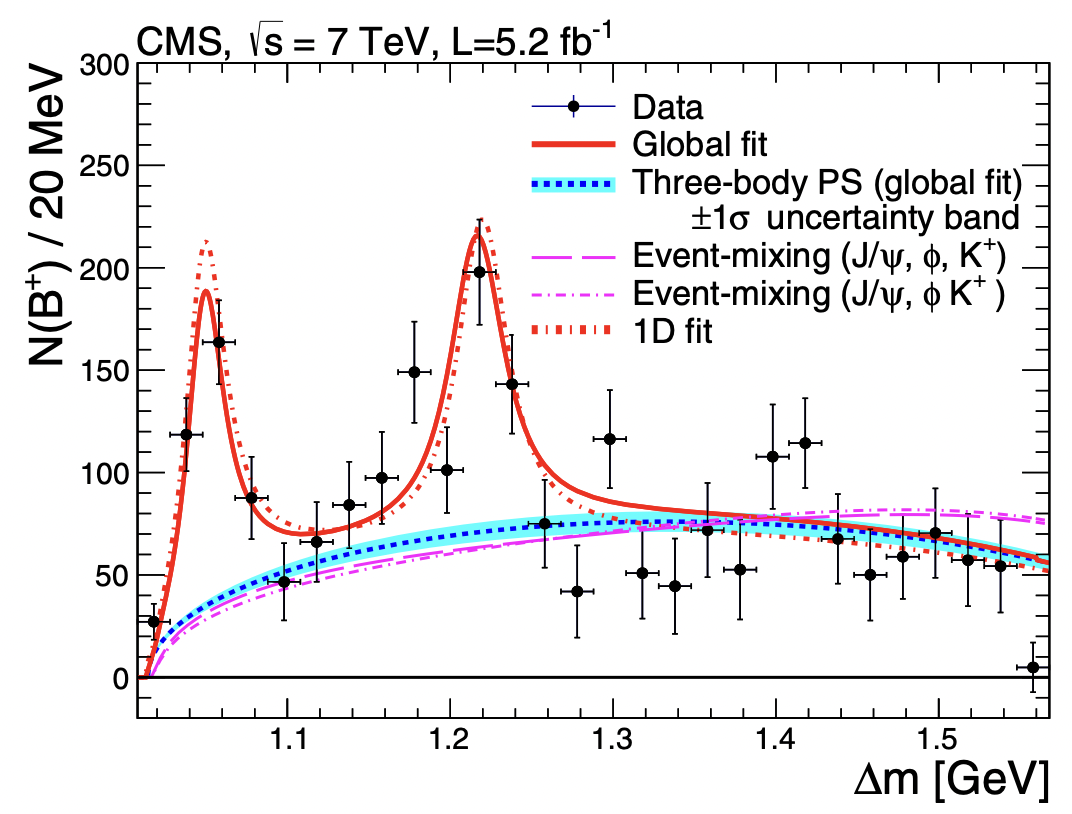}  
\caption{The fitted $Y(4140)$ distributions from two collaborations. 
\textbf{Left} (CDF): The mass difference ($\Delta m = m(\mu^+\mu^-K^+K^-)-m(\mu^+\mu^-)$) in the $B^{+}$ mass window~\cite{CDF2009}; 
\textbf{Right} (CMS): The number of $B^{+} \rightarrow J/\psi \phi K^{+}$ candidates as a function of $\Delta m = m(\mu^+\mu^-K^+K^-)-m(\mu^+\mu^-)$~\cite{CMS2014}.
}
\label{refplot}
\end{center}
\end{figure}

As an example, in~2009, the~$Y(4140)$ structure was discovered by the CDF collaboration with a local significance of 5.3$\sigma$, using a data sample corresponding to an integrated luminosity of 2.7 $fb^{-1}$~\cite{CDF2009}. The~fitted distribution is shown in Figure~\ref{refplot} (left), and~Table~\ref{refpara} lists the published fit parameters. CMS found that its global significance was 4.3$\sigma$ based on a toy Monte Carlo (MC) computation including the 'look-elsewhere-effect'.
Later in 2011, CDF updated their result with more data and calculated its global significance to be greater than 5$\sigma$~\cite{CDF2011}. 
In 2014, the~CMS experiment provided the first confirmation for the existence of $Y(4140)$ with very high significance, i.e.,~about 7.6$\sigma$~\cite{CMS2014}. CMS's fitted plot is shown in Figure~\ref{refplot} (right), and~some parameters are given in Table~\ref{refpara}. The~corresponding global significance was rather difficult to compute for such a rare fluctuation hypothesis. For~a Gaussian distribution, the~integral from 7.6$\sigma$ to positive infinity is $1.5\times10^{-14}$, which means one would need an enormous collection of $1.5\times10^{14}$ toy experiments to have, on~average, one toy instance which fluctuated above 7.6$\sigma$. 
Thus, an enormous %EE: check meaning retained
 number of toy experiments are required by the standard counting method, a~number far beyond any plausible computing~resources.

Here, we develop a new extrapolation method to evaluate large global significances without the need for enormous toy samples, and~verify it by comparing with the conventional method. Also, we compare it to the G-V method~\cite{Trial} in Section~\ref{globalcms}, which also allows the calculation of global significance with relatively small numbers of MC samples. From~the comparison, we conclude that the new method is easier to implement and interpret and gives similar results to those obtained from the G-V~method. \footnote {We became aware of a relevant paper~\cite{NIMA} after publication, thus a comparison with it cannot be done. }

\unskip

\vspace{-6pt} 

\begin{table}[H] 
\centering
\caption{The recap of some parameters from the CDF and CMS collaboration~\cite{CDF2009,CMS2014}.\label{refpara}}
\newcolumntype{C}{>{\centering\arraybackslash}X}
\begin{tabularx}{\textwidth}{CCCC}
\toprule
\textbf{Collaboration} & \boldmath{$B^+$} \textbf{Signal Events} & \textbf{Mass of Y(4140) (MeV/c$^2$)} & \textbf{Width of Y(4140) (MeV/c$^2$)} \\
\midrule
CDF & $75 \pm 10$ & $4143.0 \pm 2.9(stat) \pm 1.2(syst)$ & $11.7^{+8.3}_{-5.0}(stat) \pm 3.7(syst)$ \\
CMS & $2320 \pm 110$ & $4148.0 \pm 2.4 (stat) \pm 6.3 (syst)$ & $28^{+15}_{-11} (stat) \pm 19 (syst)$\\
\bottomrule
\end{tabularx}
\end{table}

\subsection{Global Significance of CDF'S $Y(4140)$: The Conventional, Extrapolation, and~ G-V Methods}
We will apply both our new extrapolation method and that of Gross--Vitells~\cite{Trial}  to calculate the global significance for CDF's $Y(4140)$ report, and~compare the two results with those obtained by the conventional method to test the validity of our new~method. 

Notice that the local significance from the CDF case is only 5.3$\sigma$~\cite{CDF2009}, which means that the number of MC samples needed to quantify such an excess by the conventional counting method is still~accessible.  

\subsubsection{Background and Signal~Components}
The Y(4140) structure was originally observed in the $J/\psi \phi$ mass spectrum in $B^{\pm}\to J/\psi \phi$$K^{\pm}$ decays produced in $p\Bar{p}$ collisions at $\sqrt{s}$ = 1.96 TeV collected by the CDF II detector~\cite{CDF2009,CDF2011}.
The fit functions used by CDF
to model their data included a Breit--Wigner function to describe the signal component:
\begin{equation}
f(m;m_{0},\Gamma)=\frac{1}{\pi}[\frac{\Gamma}{(m-m_{0})^2+\Gamma^2}]
\label{bwfun}
\end{equation}
where $m$ is the spectrum's variable mass, and~$m_0$ and $\Gamma$ are the resonance's mass and width.
The background component is described by the three-body phase-space function~\cite{three}:
\begin{equation}
\begin{gathered}
p(m)=\frac{\frac{1}{m^{2}}\lambda^{\frac{1}{2}}(m,m_{J/\psi},m_{\phi})\lambda^{\frac{1}{2}}(m_{B},m_{K},m)}{\int\frac{1}{m^{2}}\lambda^{\frac{1}{2}}(m,m_{J/\psi},m_{\phi})\lambda^{\frac{1}{2}}(m_{B},m_{K},m)\mathrm {d}t^{2}}, \\
\lambda^{\frac{1}{2}}(a,b,c)=\sqrt{a^4+b^4+c^4-2a^2b^2-2b^2c^2-2a^2c^2}, \\
(m_{J/\psi}+m_{\phi}) \leq m \leq (m_{B}-m_{K}),
\label{bkgfun}
\end{gathered}
\end{equation}
where $m_{J/\psi}$, $m_{\phi}$, $m_{B}$, and~$m_{K}$ are the masses of $J/\psi$, $\phi$, $B^{\pm}$, and~$K^{\pm}$, respectively. Here, $m$ is the mass of the $J/\psi \phi$ system, and~$p(m)$ is the probability distribution of $m$.

\subsubsection{Generating and Fitting Simulated~Events}
We only investigate candidates below 4.665 GeV because of possible background from $B^0_s \rightarrow \psi(2S)\phi \rightarrow J/\psi\pi^+\pi^-\phi$ at higher values. 
Therefore, in $B^{\pm}\to J/\psi \phi$$K^{\pm}$ decays, the~mass of the $J/\psi \phi$ ranges from 4.116 ($m_{J/\psi}+m_{\phi}$) to 4.665 
GeV~\cite{CDF2009,CDF2011,CMS2014}, which is the search range for the new particle. 
We use the three-body phase-space function, Equation~(\ref{bkgfun}), to~generate simulated background-only samples with a variable number of events with a mean of 75 and sigma of 10, according to the numbers reported by CDF~\cite{CDF2009}. It is computationally inefficient to work with the  unbinned sample when a sufficiently finely binned treatment will yield a good approximation. 
Thus, our first task was to determine a sufficiently small bin size by generating a small toy sample and comparing binned and unbinned results. For~the unbinned fit, we minimize the negative log-likelihood function summed over terms for each individual event, while the binned fit has likelihood terms for a relatively small number of narrow bins spanning the range of [4.116,4.665] GeV, so that events in the same bin will be processed and fitted together. The~unbinned fit will be more accurate than the binned fit, but~it is computationally much more intensive than the binned~fit.

To determine an acceptable bin size, we define the fitting procedure for a given pseudo-experiment as follows. We conduct the log-likelihood fits for the simulated samples for the null and signal hypotheses to obtain $L0$ and $L1$. For~the signal-hypothesis fit, we set up 15 loops to scan over the range of initial fit values of the parameters in the signal function. The~initial values of the signal mass are in the range [4.2,4.6] GeV (step size is 0.1 GeV), and~the signal width uses 0.005, 0.015, and~0.05 GeV. After~going through 15 loops to scan the mass-width parameter space, we find the maximum signal log-likelihood $L1$. By~computing the final $2 \times (L0-L1)$, we obtain the maximum value of the log-likelihood ratios for a given~pseudo-experiment.

\begin{figure}[H]
\begin{center}
\includegraphics[width=0.95\textwidth]{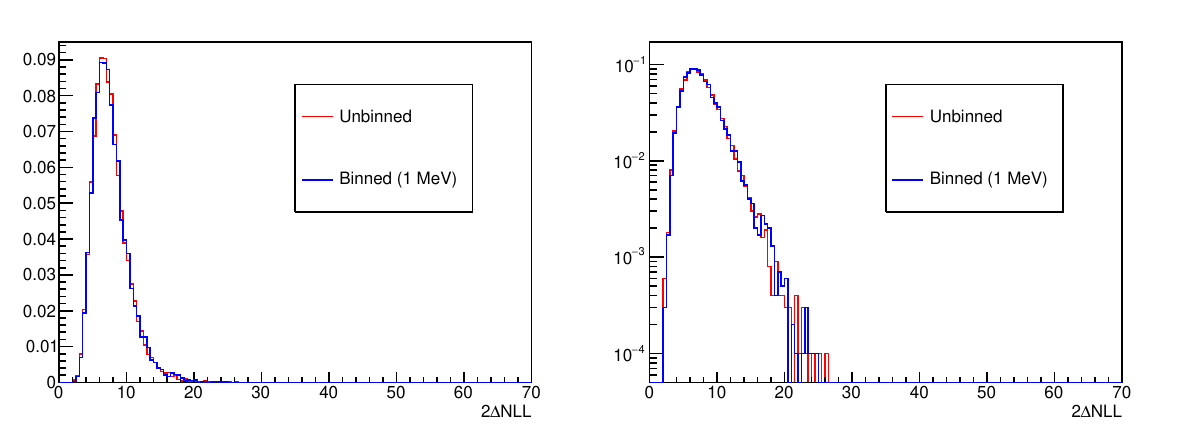}
\caption{The null-signal likelihood ratio distribution for toy MC computed for 1 MeV binned (blue) and unbinned (red) data. \textbf{Left}: linear scale; \textbf{right}: logarithmic scale.
}
\label{cdf1MeV}
\end{center}
\end{figure}

With a bin width of 1 MeV, the~shape of the log-likelihood difference distribution %EE: check meaning retained
is very similar to the unbinned data, as~can be seen in Figure~\ref{cdf1MeV}. As~the two results are very similar, we choose binned log-likelihood fits with a 1~MeV bin to save time and increase~efficiency.

\unskip

\subsubsection{The Conventional Method---Direct~Counting}
\begin{figure}[H]
\begin{center}
\includegraphics[width=0.7\textwidth]{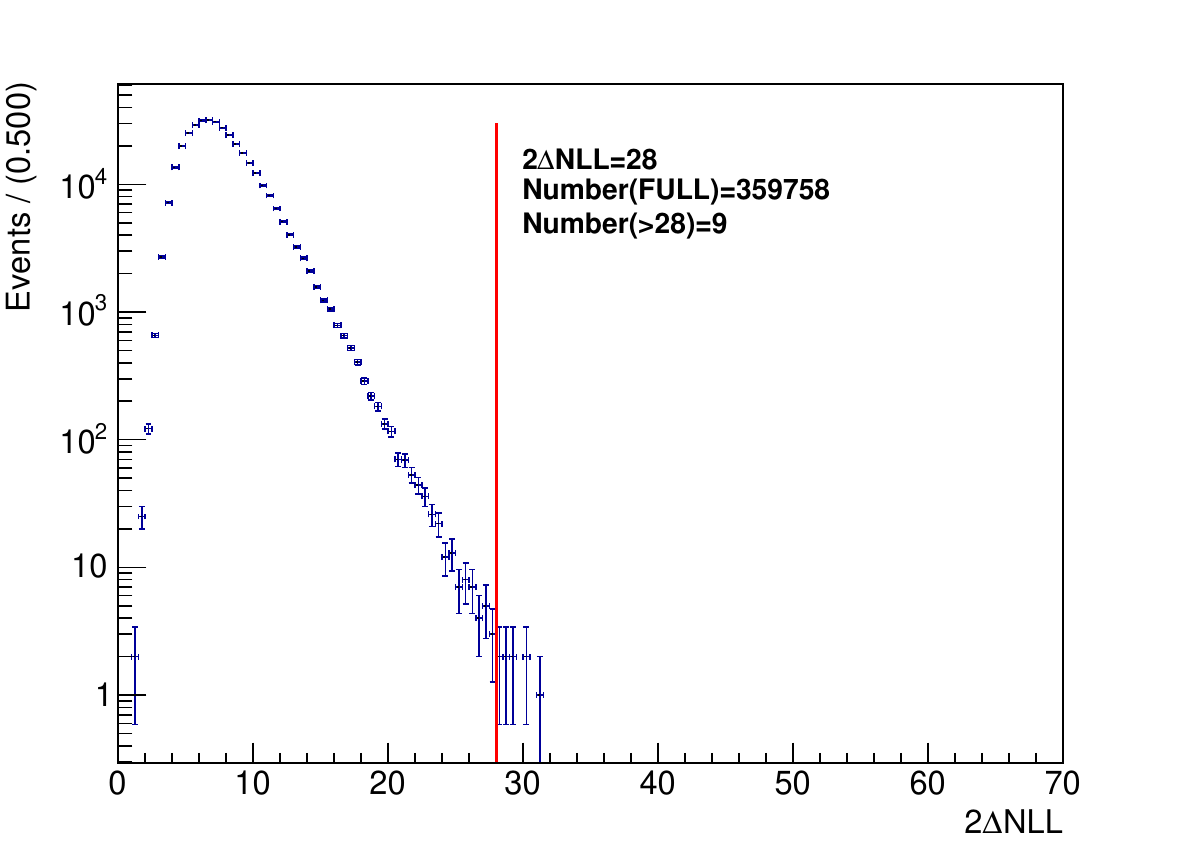}
\caption{The distribution of the binned
log-likelihood ratios of the 359,758 toy experiments simulating CDF's Y(4140) result.
}
\label{cdfhist}
\end{center}
\end{figure}

In order to demonstrate our extrapolation method, we generate a fairly large sample of toy MC, but~one which is still vastly smaller than would be needed for the conventional method. We submitted 359,758 jobs to a computer cluster, with~each job representing one toy experiment mimicking CDF's Y(4140) sample, and~thereby one log-likelihood ratio. 
We thereby obtain 359,758 simulated experiments. Figure~\ref{cdfhist} shows a histogram of the distribution of all the log-likelihood ratio values for this sample of~pseudo-experiments. 

With a local significance of 5.3$\sigma$, the~reference level of the likelihood ratio should be 28. 
We can directly count that the number of toy experiments which had a likelihood ratio beyond 28 is 9. With~the total number of experiments being 359,758, the~\emph{p}-value is $9/359,758$, or~$2.50\times10^{-5}$, for~a global significance of 4.1$\sigma$.

\subsubsection{The Extrapolation~Method}
\label{cdfextra}
\begin{figure}[H]
\begin{center}
\includegraphics[width=0.7\textwidth]{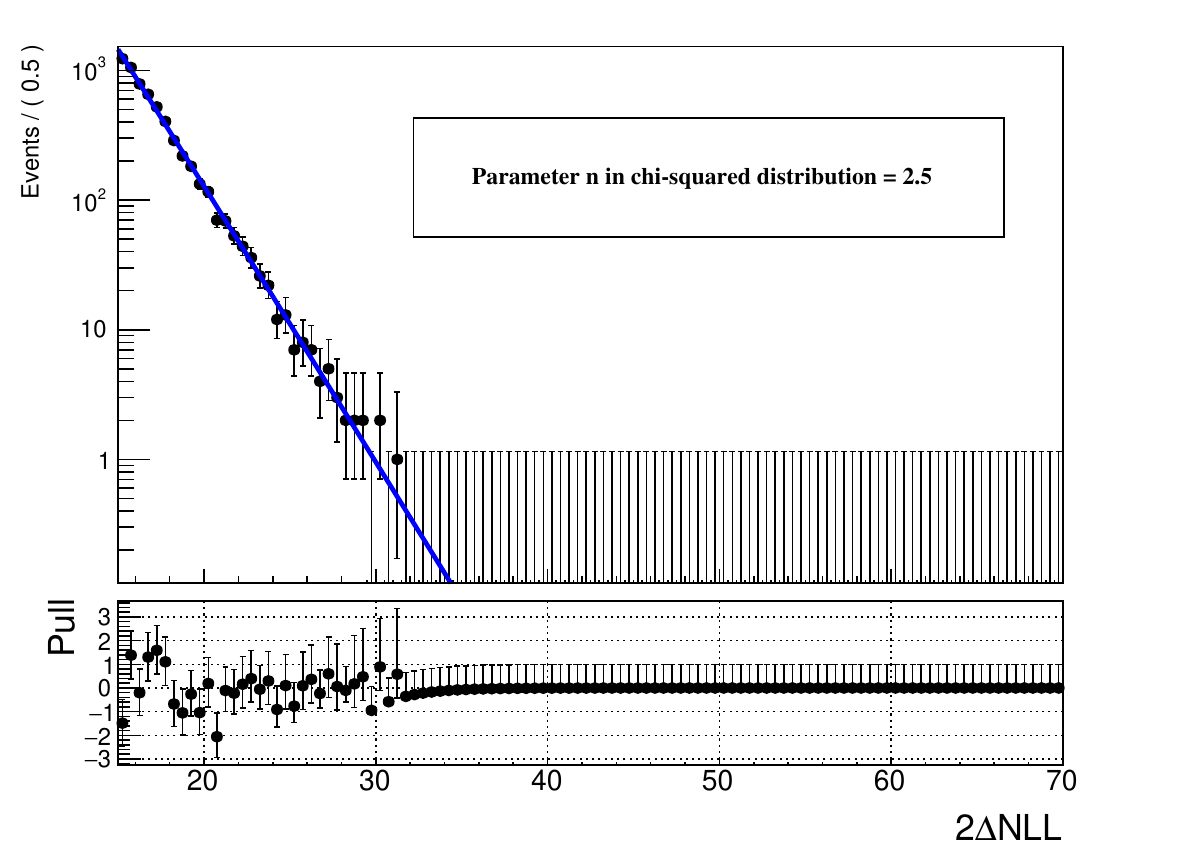}  
\caption{The tail of the likelihood ratio distribution on a log scale and fit with a $\chi^2$ function for the CDF case. The~reduced $\chi^2$ value of 0.69 indicates reasonably good fit quality.
}
\label{cdffit}
\end{center}
\end{figure}

The strategy behind our extrapolation method is to exploit the fact that the tail of the log-likelihood ratio distribution---which drives the significances---is well approximated by a $\chi^2$-distribution. The~first step is to determine where to begin the tail we want to model. After~some trial and error, we found that using a tail  above ratios of 15 worked well. Thus, we divided all the simulated data into two ratio ranges: [0,15] and [15,70]. 
We separated the latter range into 110 bins, and~used the $\chi^2$ distribution as the PDF to fit the latter range. Its probability density function is
\begin{equation}
f(x) = 
\begin{cases}
    \frac{1}{2^{n/2}\varGamma(n/2)}x^{\frac{n}{2}-1}e^{-x/2} & \text{if } x>0 \\
    0 & \text{if } x\leq0,
\label{chisq}
\end{cases}
\end{equation}
where $n$ is the number of degrees of freedom.
The likelihood ratio distribution on a log scale and fit with a $\chi^2$ function is shown in Figure~\ref{cdffit}.

The $\chi^2$-distribution well describes the high tail of the likelihood ratio in Figure~\ref{cdffit}. 
So, we can easily project to the expected number of toys that will exceed some given likelihood ratio because
the tail of the $\chi^2$ distribution at large values is linear on a log scale.
The likelihood ratio observed in the CDF experiment was 28, and~from
integrating the fitted $\chi^2$ function,
the expected number of toy experiments above that value was 10.4.
On the other hand, the~corresponding total toy sample is broken into two pieces: those  below a ratio of 15, and~those above.
We directly count the former to be  353,780 toys, and the latter is obtained by integrating the fitted $\chi^2$ function above 15, for~a rounded value of 5,978.
Thus, the total toy sample is 359,758, and~the \emph{p}-value is $10.4/359,758$, which is about $2.89\times10^{-5}$. Assuming the fluctuations are Gaussian distributed, this corresponds to a global significance of 4.0$\sigma$.

Thus, we find that the global significance evaluated by our extrapolation method is very close to that calculated by the conventional counting method. Notice that our goal is to use the extrapolation method to estimate the global significance without aiming for precision to several decimal places, so a difference of 0.1$\sigma$ is within the margin of uncertainty we allow. Therefore, we successfully verified the validity of this simple extrapolation~method.

\subsubsection{The G-V~Method}
\label{cdfgv}
As an additional cross-check, we use the G-V method to verify the effectiveness of the extrapolation~method.

The G-V method's strategy involves identifying the frequency of fluctuations beyond the standard defined by local significance, which represents the probability of excesses occurring anywhere in the whole mass range, rather than just at the specific location of a particular fluctuation. The formula is given by
\begin{equation}
p_{global}=p_{local}+<N(c)>
\label{gvpreli}
\end{equation}
where $N(c)$ denotes the number of 'upcrossings' of the level $c$ determined by the likelihood ratio in local significance. However, $<N(c)>$ is usually difficult to calculate for large values of $c$, so it was proposed to estimate the expected number of upcrossings of likelihood ratios at a low reference level $c_{0}$ using a small set of background-only Monte Carlo simulations. Therefore, Equation~(\ref{gvpreli}) becomes
\begin{equation}
p_{global}=p_{local}+<N(c_{0})> \times (c/c_{0})^{\frac{s-1}{2}} \times e^{-(c-c_{0})/2}
\label{gvmethod}
\end{equation}
where $s$ are degrees of freedom of the $\chi^2$ distribution, which the likelihood ratio~follows.

\begin{figure}[H]
\begin{center}
\includegraphics[width=0.7\textwidth]{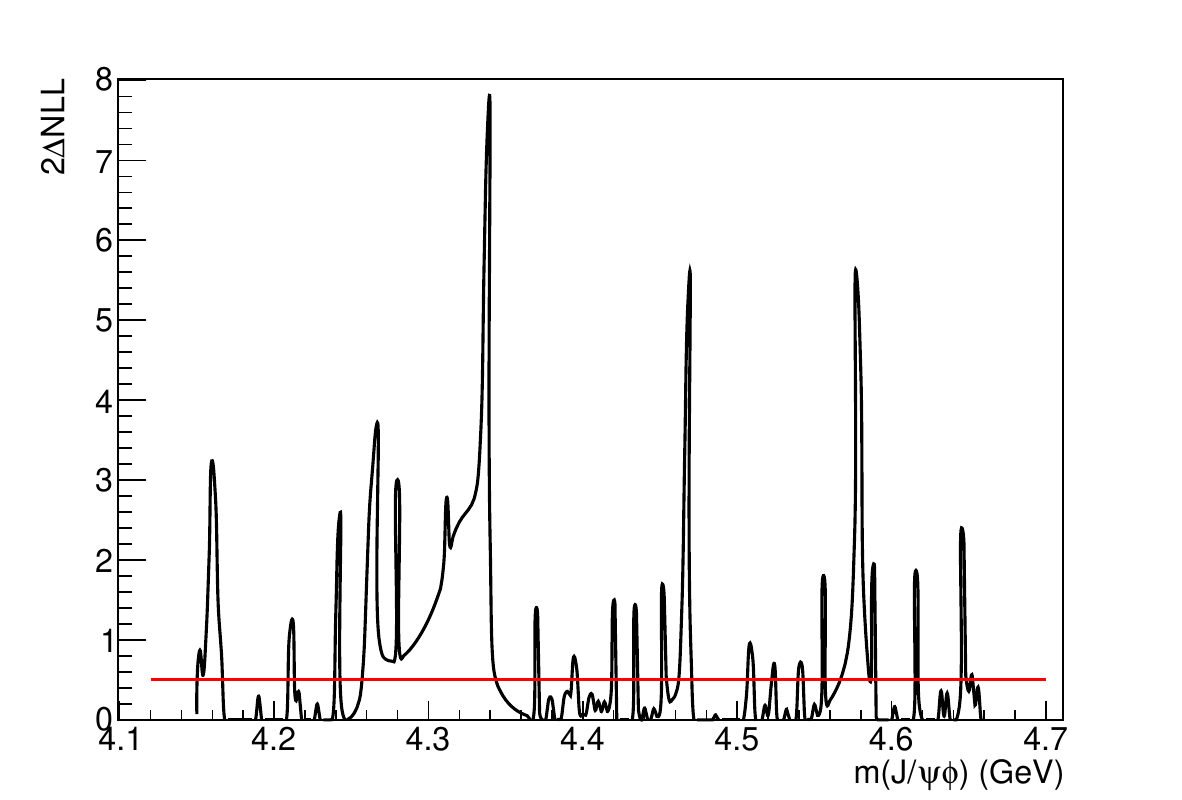}  
\caption{The likelihood ratio fluctuation curve obtained by fixing the signal mass parameter in the fit to the CDF $Y(4140)$ toy Monte Carlo sample. The~horizontal axis represents the fixed signal mass, while the vertical axis represents the obtained likelihood ratios. The~red solid line shows the $c_{0}$ reference level, which is crossed about 19 times with the  test of statistics ($N(c_{0})$=19). 
}
\label{cdfscan}
\end{center}
\end{figure}

CDF reported the local significance of $Y(4140)$ to be 5.3$\sigma$, corresponding to \emph{p}-value $p_{local}=5.79 \times 10^{-8}$ and likelihood ratio $c=28$. We have $s=1$ (1 channel), so $c_{0}=0.5$~\cite{Trial}. We performed 20 such pseudo-experiments in the mass range [4.116,4.665] GeV; one likelihood ratio distribution with $N(c_{0})=19$ is shown in Figure~\ref{cdfscan} and the average value, $<N(c_{0})>$, for~20 experiments is found to be $15.55 \pm 4.41$. Along with other inputs, we obtain the global \emph{p}-value $=1.67 \times 10^{-5}$ from Equation~(\ref{gvmethod}). The~corresponding global significance is 4.1$\sigma$. 

The CDF experiment reported the global significance of $Y(4140)$ to be 4.3$\sigma$. Compared with our results, for~direct counting, extrapolation, and~the G-V method, the~global significances are  4.1$\sigma$, 4.0$\sigma$, and~4.1$\sigma$, respectively.

\subsection{Global Significance of CMS'S $Y(4140)$ with the Extrapolation Method and the G-V Method}
\label{globalcms}

As mentioned in Section~\ref{limitmc}, the~number of MC samples ($1.5 \times 10^{14}$) required to calculate the global significance for CMS's $Y(4140)$~\cite{CMS2014} is so large that it would be difficult to realize, so the conventional direct-counting method cannot be~applied. 

But in the previous section, using  CDF's $Y(4140)$, the~new extrapolation method and the G-V method were successfully validated, so they can be used to calculate the global significance with a smaller toy sample. So, in this section, we will use the extrapolation method and the G-V method to evaluate the global significance for the CMS's $Y(4140)$, which has never been calculated before. The~two methods can be compared to each~other.

There are in total $2320 \pm 110$ events in the mass range [4.116,4.665] GeV in the CMS $J/\psi \phi$ mass spectrum~\cite{CMS2014}. We need to generate around 2320 background-only events for each simulated experiment, according to phase-space, Equation~(\ref{bkgfun}). For~a local significance of 7.6$\sigma$ observed in the CMS case, the likelihood ratio is 58. Unfortunately, as~expected, after~272,443 simulated experiments generated, none of them have a likelihood ratio above 58, indicating that the conventional method is practically inapplicable in this~case.

\begin{figure}[H]
\begin{center}
\includegraphics[width=0.7\textwidth]{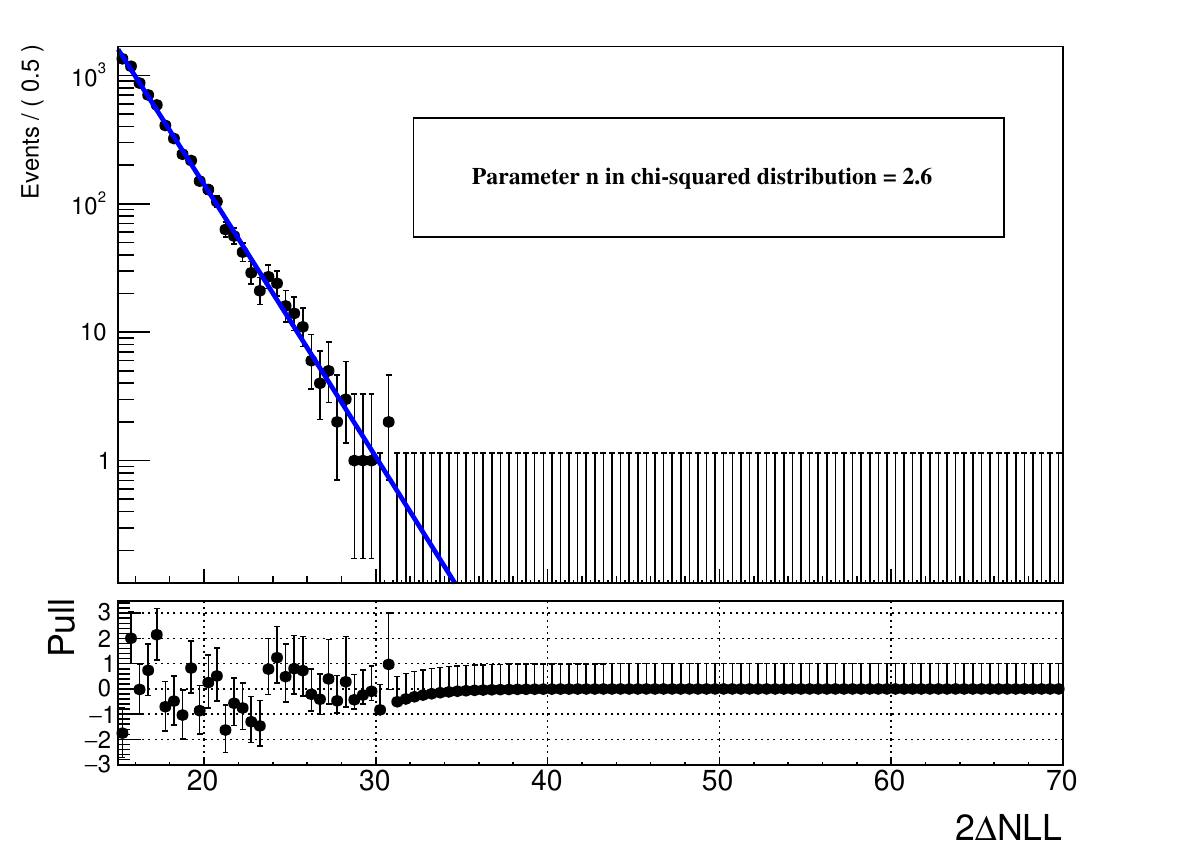}  
\caption{The likelihood ratio distribution on a log scale fit with a $\chi^2$ function for CMS's $Y(4140)$ observation. The~reduced $\chi^2$ value of 0.98 indicates reasonably good fit quality.
}
\label{cmsfit}
\end{center}
\end{figure}

However, with~the extrapolation method, we can still determine the number of toy experiments expected above a likelihood ratio of 58 by the integrals of  Equation~(\ref{chisq}). After~the fit (shown in Figure~\ref{cmsfit}), we calculate the extrapolated number of toys above 58, which should be around $4.30 \times 10^{-6}$. 
And the total number of toys is obtained by directly counting from 0 to 15 plus a integral from 15 to 70, which is 272443. 
We then compute the \emph{p}-value as $4.30 \times 10^{-6}/272,443 = 1.58 \times 10^{-11}$, and~the corresponding global significance is 6.6$\sigma$.

\begin{figure}[H]
\begin{center}
\includegraphics[width=0.7\textwidth]{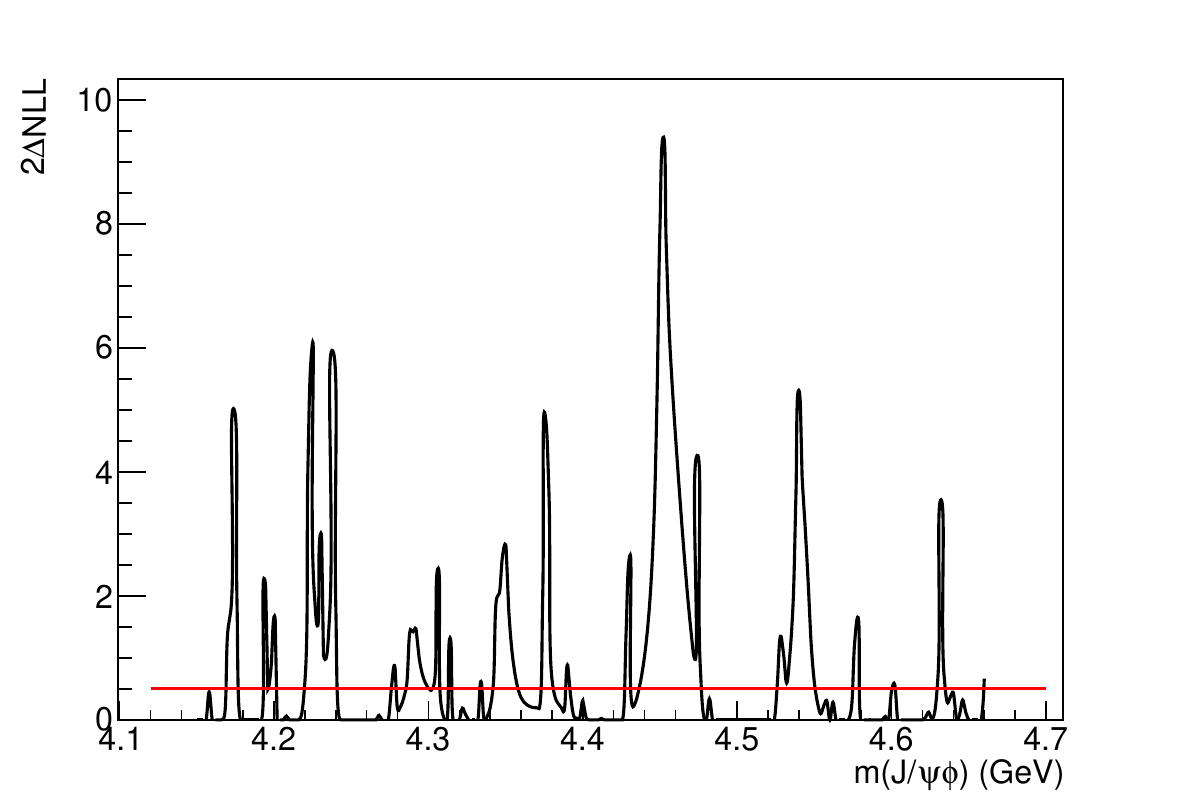}  
\caption{The likelihood ratio fluctuation curve obtained by fixing the signal mass parameter in the fit to the CMS $Y(4140)$ toy Monte Carlo sample. The~red solid lines shows the $c_{0}$ reference level, which crosses about 19 times with the test of statistics.
}
\label{cmsscan}
\end{center}
\end{figure}

For the G-V method, a~test of statistics $2 \times (L0-L1)$ is shown in Figure~\ref{cmsscan}. 
The highlights~are as follows: 
\begin{itemize}
    \item With a local significance of 7.6$\sigma$, $p_{local}$ should be $1.48 \times 10^{-14}$;
    \item Like the CDF case, $s=1$ (1 channel), and~$c_{0}=0.5$; 
    \item With 20 pseudo-experiments in the mass range [4.116,4.665] GeV, the~value of $<N(c_{0})>$ is found to be $18.80 \pm 2.75$. 
\end{itemize}

With Equation~(\ref{gvmethod}) and the above parameters, $p_{global}$ is calculated to be $6.16 \times 10^{-12}$, which give us a global significance of 6.8$\sigma$. 

\vspace{-9pt} 

For an analyzer, a~global significance level of 6.8$\sigma$ is considered statistically significant, as~it exceeds the commonly accepted threshold of 5$\sigma$. Then, a difference of 0.2$\sigma$ between 6.8$\sigma$ (from the G-V method) and 6.6$\sigma$ (from extrapolation) may not be considered substantial. And~their corresponding \emph{p}-values differ by an order of magnitude, making error calculation challenging. Anyway, it is generally unnecessary to focus on small differences at this level (6.8 vs 6.6), unless~they are close to critical thresholds such as 3$\sigma$ or 5$\sigma$. On~the other hand, 5$\sigma$ is low enough to use the conventional method to obtain the more accurate~result. 

\subsection{Global Significance of ATLAS'S $ \chi_b(3P)$ with the Extrapolation Method and the G-V Method}
\label{globalatlas}

\begin{figure}[H]
\begin{center}
\includegraphics[width=0.67\textwidth]{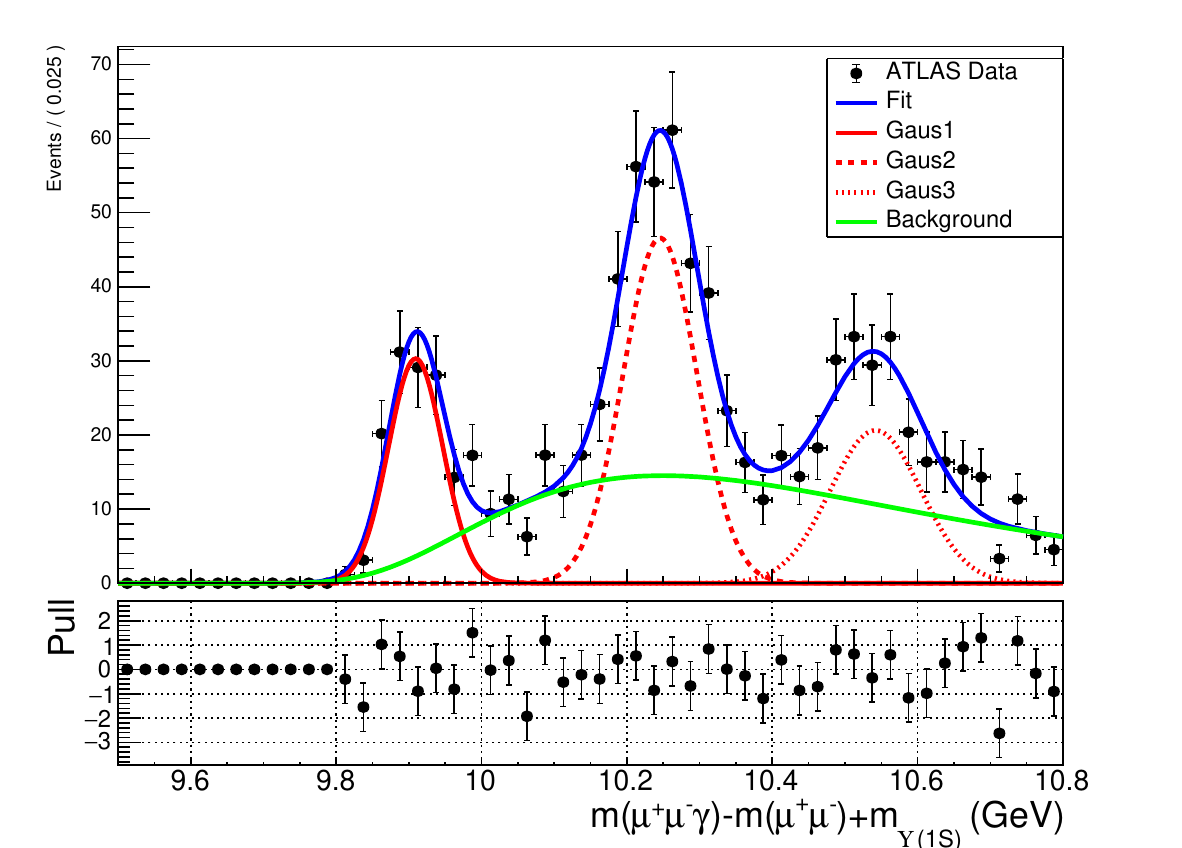}  
\caption{Reproducing ATLAS experiment's $m(\mu\mu\gamma)-m(\mu\mu)+m(\Upsilon(1S))$ invariant mass spectrum with simulated events. The~third peak from the left corresponds to the $\chi_b(3P)$ observed in 2012 by ATLAS. 
}
\label{atlasdata}
\end{center}
\end{figure}

To further demonstrate the versatility of the new method across different datasets, we applied it to a resonance named $\chi_b(3P)$, reported by the ATLAS collaboration in 2012~\cite{ATLAS}. This resonance represents the first new particle discovered at the LHC among its four major experiments. Using data points and fitting functions extracted from the ATLAS results, we reproduced the ATLAS data fit, as~shown in Figure~\ref{atlasdata}. 
%%%%Please notice the figure has been updated!!

\vspace{-6pt} 
In the ATLAS $m(\mu\mu\gamma)m(\mu\mu)+m(\Upsilon(1S))$  mass spectrum, a~total of 873 $\pm$ 71 events were collected within the mass range [9.5,10.8] GeV. Assuming the first two Gaussian peaks are also background for the search of $\chi_b(3P)$, we generated approximately 873 background-only events in each simulated experiment. Given the local significance of $6\sigma$ reported by ATLAS, the~corresponding likelihood ratio is 36. To~calculate the global significance, 60219 pseudo-experiments are~generated.

\begin{figure}[H]
\begin{center}
\includegraphics[width=0.67\textwidth]{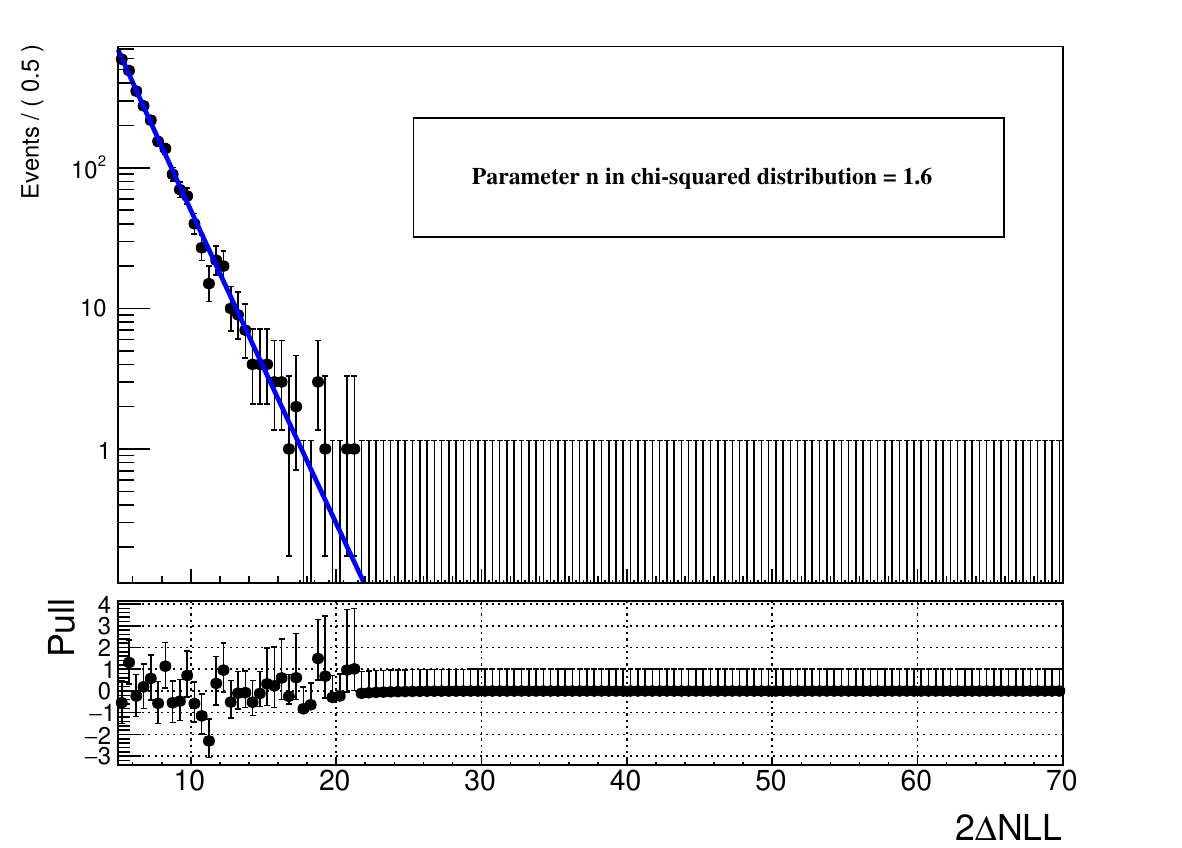}
\caption{The likelihood ratio distribution on a log scale fit with a $\chi^2$ function for ATLAS’s
$\chi_b(3P)$ observation. The~reduced $\chi^2$ value of 0.71 indicates reasonably good fit quality.
}
\label{atlasgs}
\end{center}
\end{figure}

For each experiment, we searched for fluctuations mimicking the $\chi_b(3P)$ signal in the mass region [10.48,10.60] GeV with a step size of 0.01 GeV. We also scanned signal widths in the range [48,77] MeV, spanning one standard deviation of the value obtained in the ATLAS analysis, with~a step size of 1 MeV. The~resulting likelihood ratio distribution, plotted on a logarithmic scale and fitted with a $\chi^2$ function, is shown in Figure~\ref{atlasgs}.
After extrapolation, we obtained a result of $(3.53 \pm 0.07) \times 10^{-04}$, corresponding to a \emph{p}-value of $5.87 \times 10^{-09}$ and a global significance of $5.7\sigma$. 

As a cross-check, we also calculated the global significance using the G-V method and obtained $5.6\sigma$. The~two results are consistent and we found that the agreement improves further with an increased number of~pseudo-experiments.
\unskip

\subsection{Correlations}
\begin{figure}[H]
\begin{center}
\includegraphics[width=0.3\textwidth]{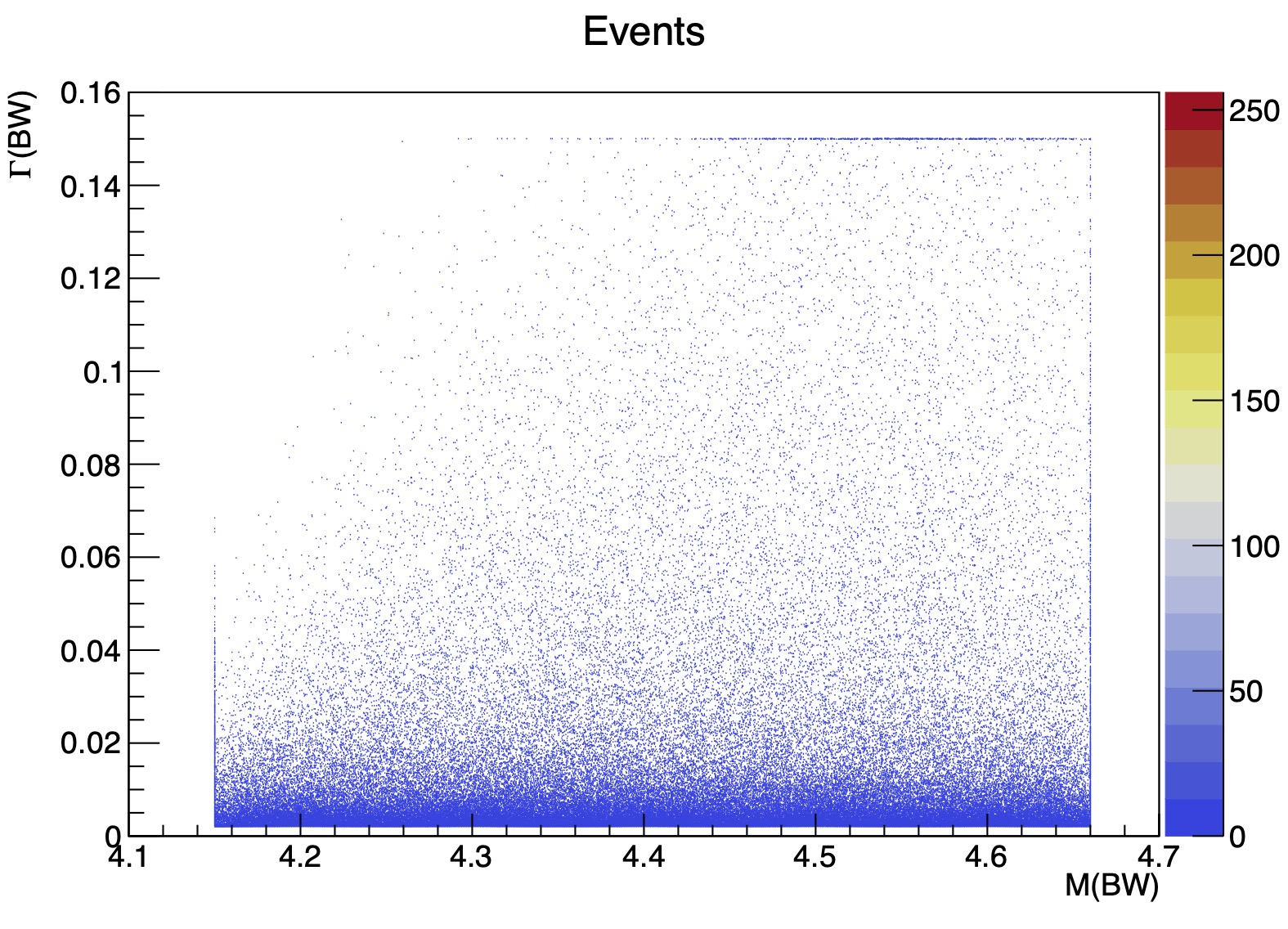}
\includegraphics[width=0.3\textwidth]{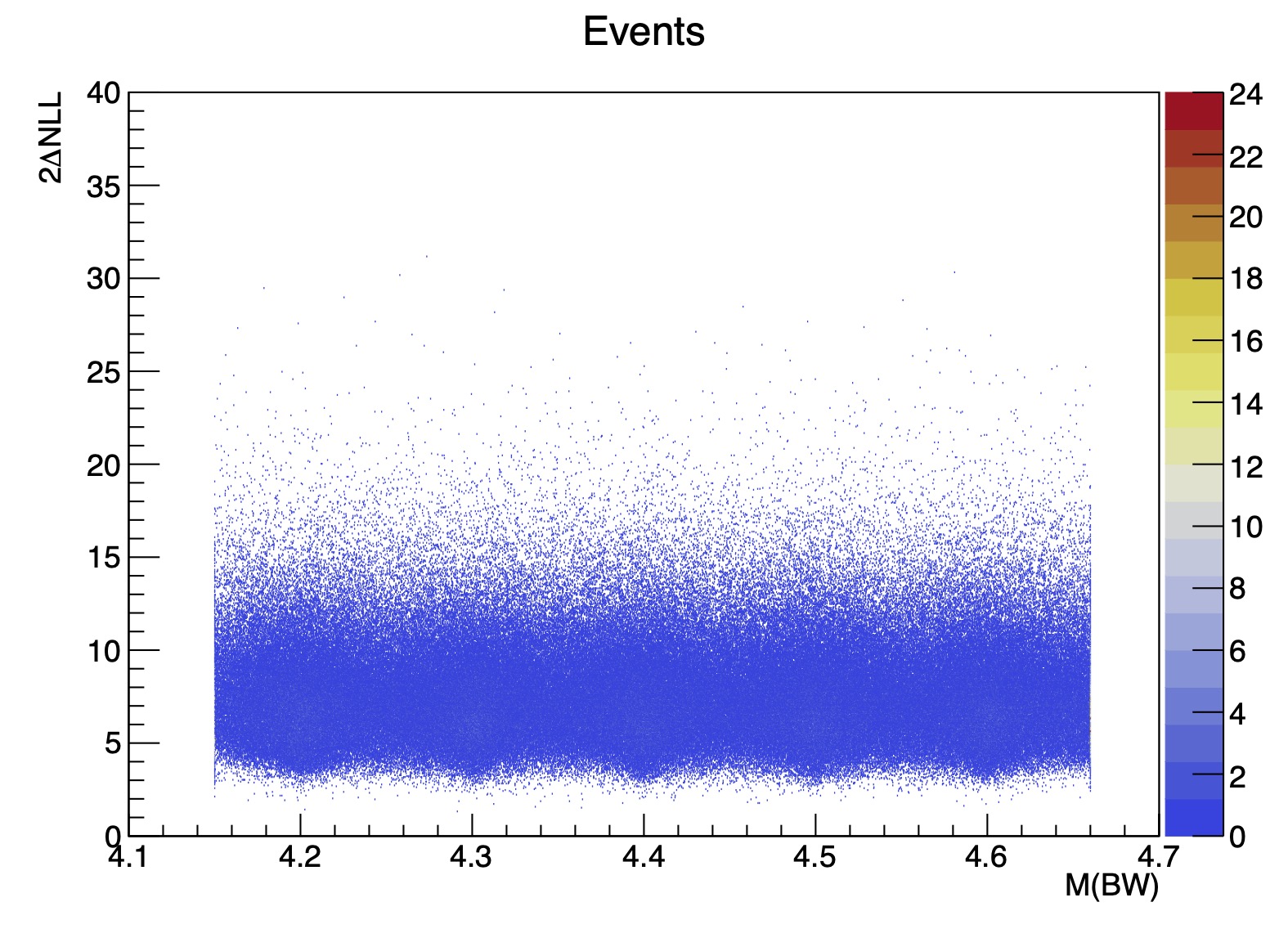}
\includegraphics[width=0.3\textwidth]{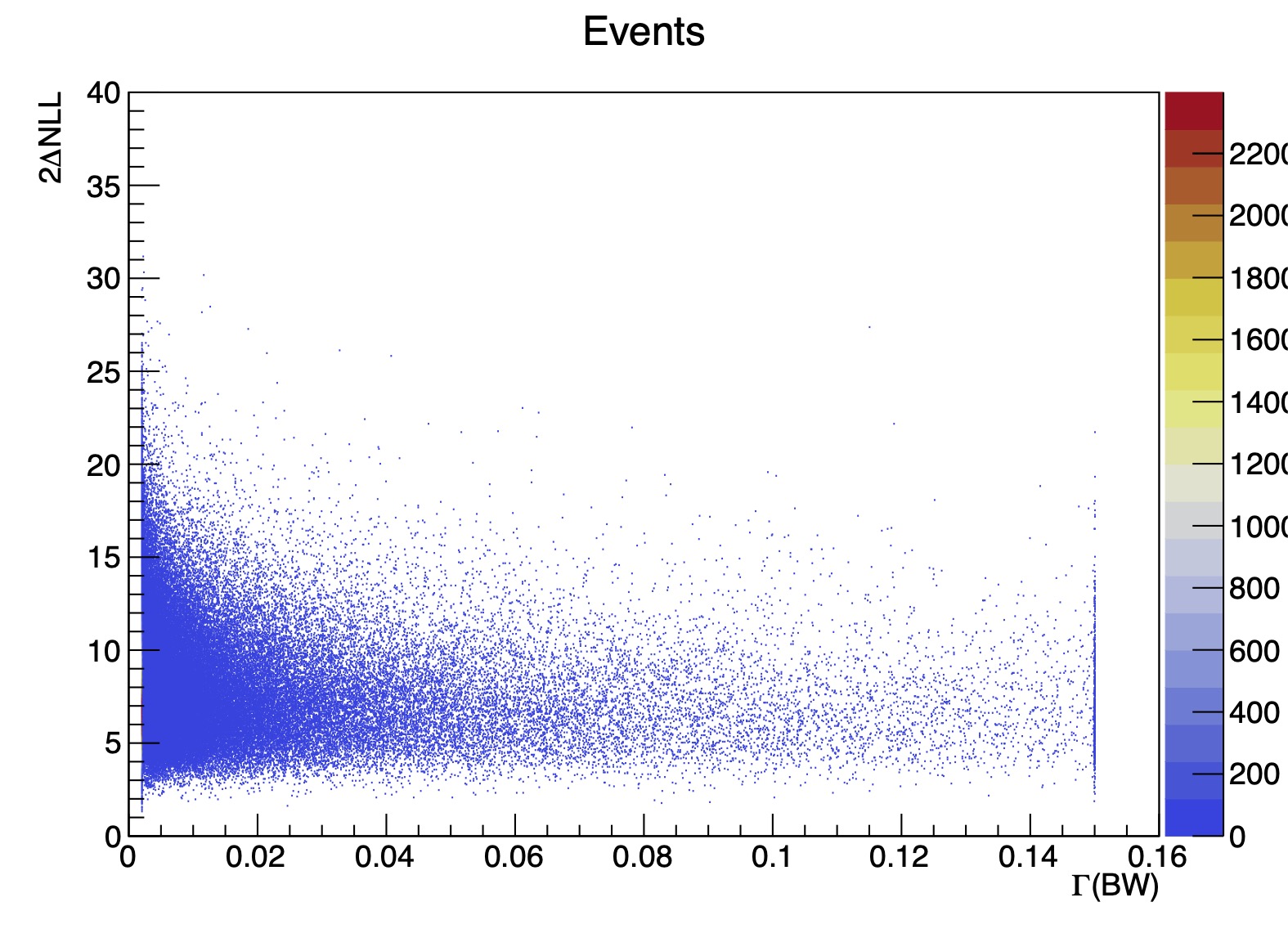}
\includegraphics[width=0.3\textwidth]{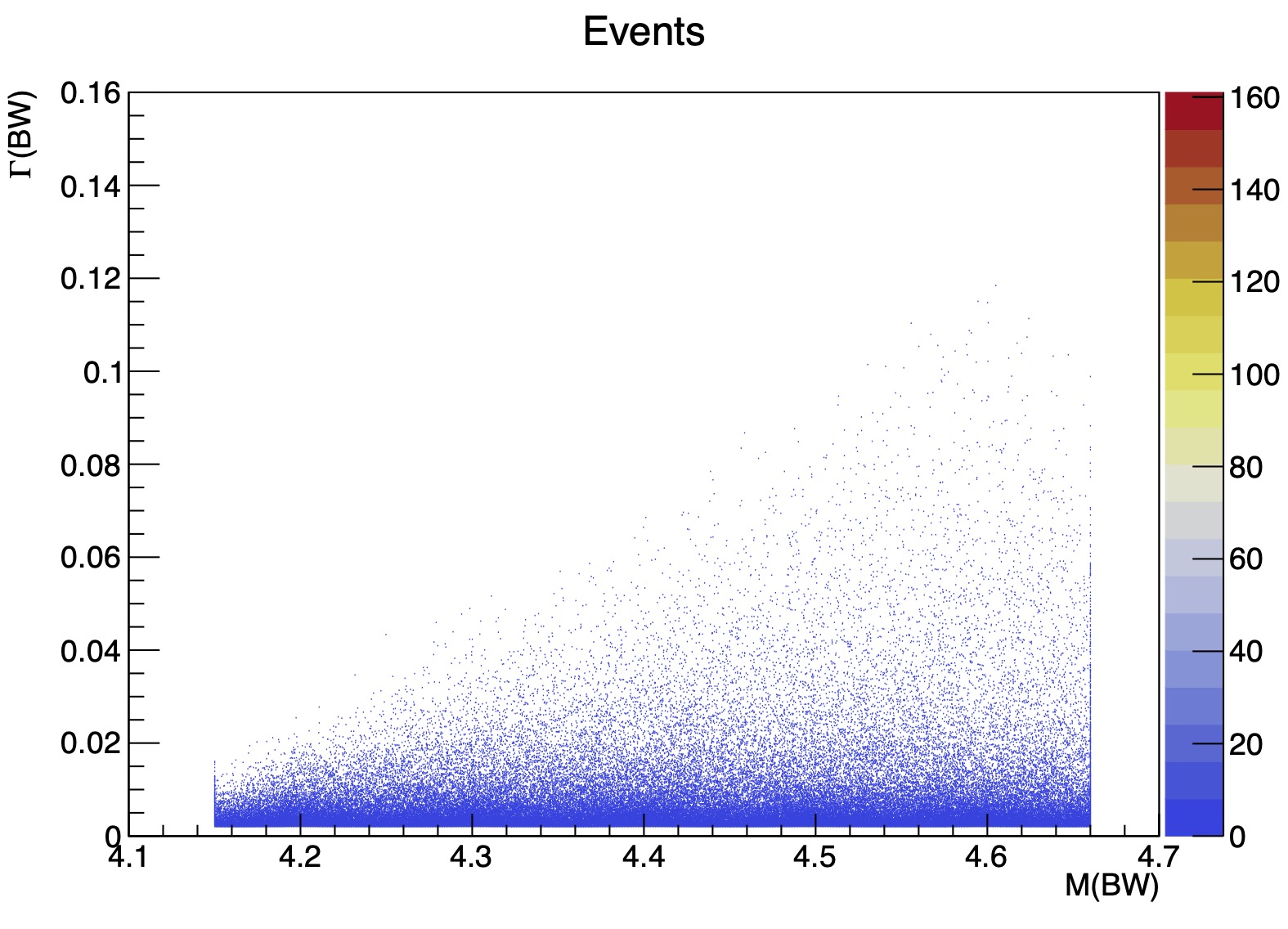}
\includegraphics[width=0.3\textwidth]{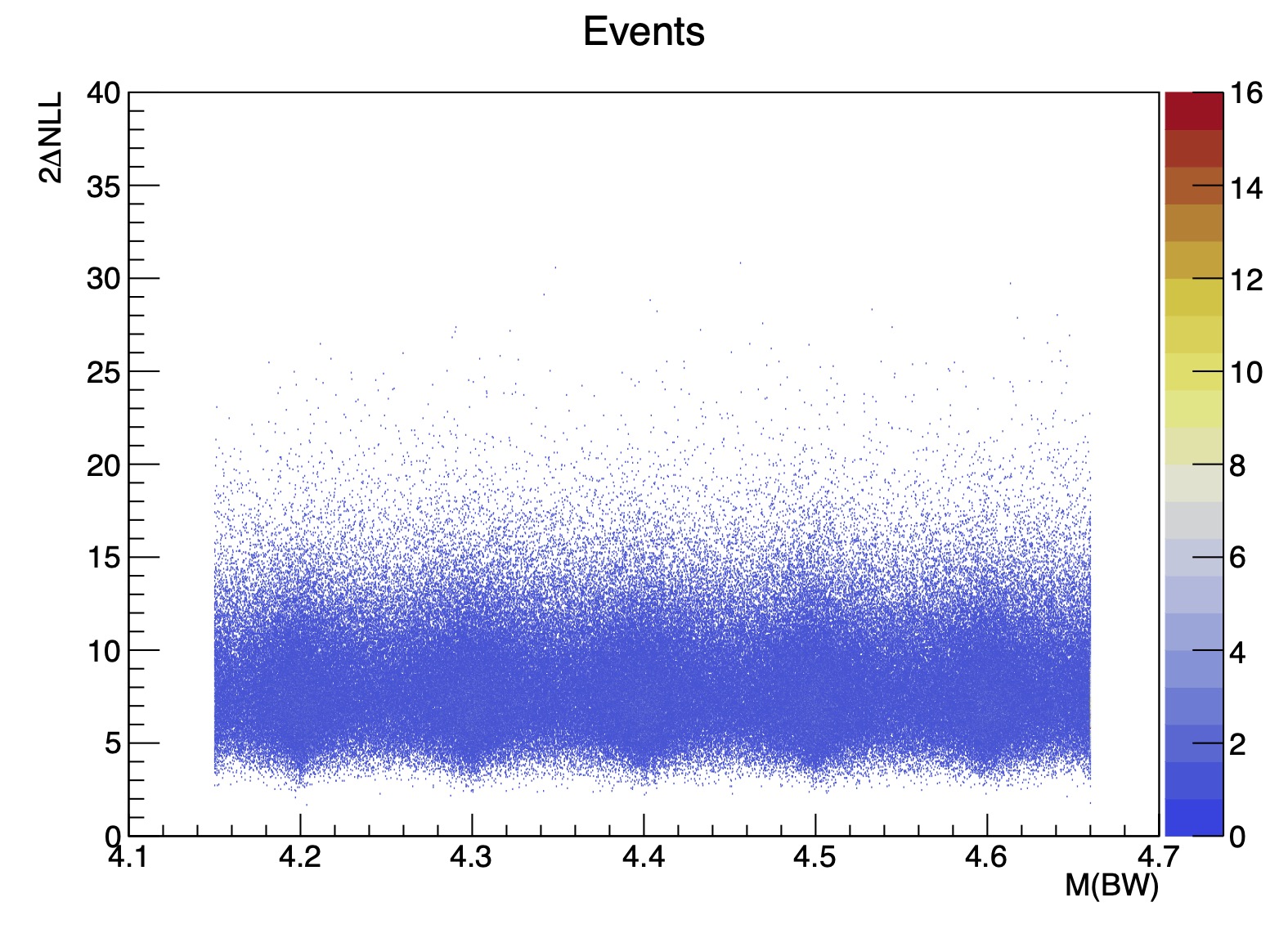}
\includegraphics[width=0.3\textwidth]{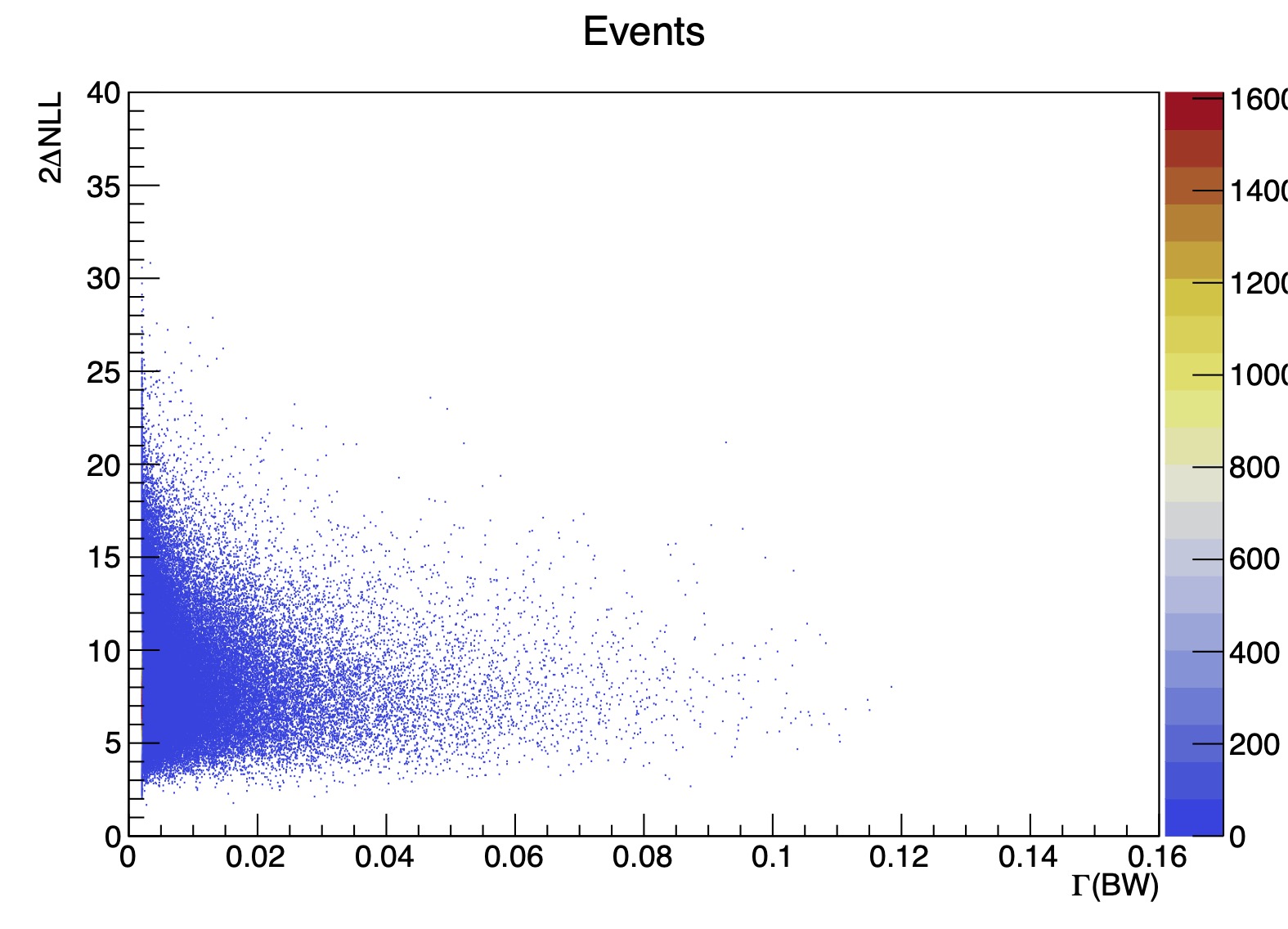}
 
\caption{
Upper (lower) \textbf{left}: signal width  {vs.} signal mass in the CDF (CMS) case, indicating a small correlation between the signal width and signal~mass.  Upper (lower) \textbf{middle}: $2 \times (L0-L1)$ {vs.} signal mass in the CDF (CMS) case, indicating the signal mass has minimal influence on the~significance.   
Upper (lower) \textbf{right}: $2\times (L0-L1)$ {vs.} signal width in CDF (CMS) case, indicating that a narrower signal width can lead to higher significance levels. }
\label{corr}
\end{center}
\end{figure}

Sometimes, we want to check if there are any correlation among the significance, signal mass, and~signal width. We produced toy MC experiments and plotted the 2D distributions for the log-likelihood ratio, signal mass, and~signal width, as~shown in Figure~\ref{corr}. 
From the right column of the figures, it is evident that a narrower signal width can lead to higher significance levels. The~middle column of Figure~\ref{corr} reveals that the signal mass has minimal influence on the significance. The~left column of Figure~\ref{corr} indicates that there is little correlation between the signal width and signal~mass. 

Overall, the~log-likelihood ratio distribution in terms of fluctuated signal mass or width towards a big fluctuation may appear randomly and uniformly anywhere in the mass region, but~often with a very narrow width. So, when we think about look-elsewhere effects, we need to notice that the width distribution is not~linear. 

\unskip

\section{Conclusions}
We demonstrate the conventional method of calculating local and global significances, and~the way to estimate the increase in significance by adding new data. 
We also developed a simple and effective extrapolation method to estimate large global significances for situations in which the standard brute-force method of generating enormously large toy MC samples is not computationally viable. 
Using the CDF and CMS reports of the $Y(4140)$ as well as the ATLAS report of the $\chi_b(3P)$ as test cases, with~results summarized in Table~\ref{finalres}, the~extrapolation method is consistent with the conventional and G-V methods.  
As a result, we conclude that the extrapolation method gives good estimation of the global significance with toy MC samples of modest sizes. And~compared to the G-V method, the~extrapolation method is easier to understand and use, and closer to the conventional~method.

\begin{table}[H] 
\caption{The global significances with different methods for the  CDF, CMS, and ATLAS~examples.\label{finalres}}
\newcolumntype{C}{>{\centering\arraybackslash}X}
\begin{tabularx}{\textwidth}{CCCC}
\toprule
\textbf{}	 & \textbf{Direct Counting}	& \textbf{Extrapolation } & \textbf{G-V Method} \\
\midrule
CDF's $Y(4140)$ & 4.1$\sigma$ & 4.0$\sigma$ & 4.1$\sigma$ \\
CMS's $Y(4140)$ & -- & 6.6$\sigma$ & 6.8$\sigma$ \\
ATLAS's $\chi_b(3P)$ & -- & 5.7$\sigma$ & 5.6$\sigma$ \\
\bottomrule
\end{tabularx}
\end{table}

%%%%%%%%%%%%%%%%%%%%%%%%%%%%%%%%%%%%%%%%%%
\vspace{6pt} 

%%%%%%%%%%%%%%%%%%%%%%%%%%%%%%%%%%%%%%%%%%

\funding{This work is partially supported by the Nanjing Normal University research start-up funding project, the Tsinghua University Initiative Scientific Research Program and Dushi Program, the Natural Science Foundation of China under Grants No. 12075123 and~No. 12061141002, and the Ministry of Science and Technology of China under Grants No. 2023YFA1605804 and No.~2024YFA1610501.}

\dataavailability{The data will be available on request.} 

\conflictsofinterest{The authors declare no conflicts of~interest.} 


\reftitle{References}


\begin{thebibliography}{999}
\bibitem{CMS_2024} CMS Collaboration. New Structures in the 
$J/\psi J/\psi$ Mass Spectrum in Proton-Proton Collisions at $\sqrt{s}=13$ TeV. 
 {\em Phys. Rev. Lett.} {\bf 2024}, {\em 132}, 111901.
 
\bibitem{CPC2024} Zhu, F.; Bauer, G.; Yi, K. Experimental Road to a Charming Family of Tetraquarks... and Beyond.  {\em Chin. Phys. Lett.} {\bf 2024}, {\em 41}, 111201.

\bibitem{SB2024} Liu, X. Four-Charm-Quark Matter from the CMS Collaboration as a witness of the development of high-precision hadron spectroscopy.  {\em Sci. Bull.} {\bf 2024}, {\em 69}, 2802-2803.

\bibitem{CDF2009} CDF Collaboration. Evidence for a Narrow Near-Threshold Structure in the $J/\psi\phi$ Mass Spectrum in $B^+\to J/\psi\phi K^+$ Decays.  {\em Phys. Rev. Lett.} {\bf 2009}, {\em 102}, 242002. 

\bibitem{IJMPA2013} Yi, K. Experimental review of structures in the $J/\psi\phi$ mass spectrum.  {\em Int. J. Mod. Phys. A} {\bf 2013}, {\em 28}, 1330020. 

\bibitem{CDF2011} CDF Collaboration. Observation of the $Y(4140)$ Structure in the $J/\psi\phi$ Mass Spectrum in $B^\pm\to J/\psi\phi K^\pm$ Decays.  {\em Mod. Phys. Lett. A} {\bf 2017}, {\em 32}, 1750139.

\bibitem{CMS2014} CMS Collaboration. Observation of a peaking structure in the $J/\psi\phi$ mass spectrum from $B^\pm\to J/\psi\phi K^\pm$ decays.  {\em Phys. Lett. B} {\bf 2014}, {\em 734}, 261-281. 

\bibitem{Trial} Gross, E.; Vitells, O. Trial factors for the look elsewhere effect in high energy physics.  {\em Eur. Phys. J. C} {\bf 2010}, {\em 70}, 525-530.

\bibitem{ATLAS} ATLAS Collaboration. Observation of a new $\chi_b$ state in radiative transitions to $\Upsilon(1S)$ and $\Upsilon(2S)$ at ATLAS. {\em Phys. Rev. Lett.} {\bf 2012}, {\em 108}, 152001.

\bibitem{Byron} Byron, P.R. \textit{Probability and Statistics in Experimental Physics}; Springer Science \& Business Media:  Berlin/Heidelberg, Germany, 1992.  % Book

\bibitem{Kelly} Yi, K.J.; Spiegel, L.; Hu, Z. A global significance evaluation method using simulated events. \emph{arXiv}  \textbf{2023}, arXiv:2310.14317.

\bibitem{CMS:2022yhl} CMS Collaboration. Observation of new structures in the $\mathrm{J}/\psi \mathrm{J}/\psi$  mass spectrum in $\mathrm{p}\mathrm{p}$ collisions at $\sqrt{s} = 13$ TeV. CMS-PAS-BPH-21-003. 2022. Available online: \url{https://cds.cern.ch/record/2815336}  (accessed on 9 July 2022). 
\bibitem{root} Brun, R.; Rademakers, F. ROOT: An object oriented data analysis framework. {\em Nucl. Inst. Meth. Phys. Res.} {\bf 1997}, {\em 389}, 81-86. 

\bibitem{hepdata} CMS Collaboration. Observation of New Structure in the $\mathrm{J}/\psi \mathrm{J}/\psi$
 Mass Spectrum in Proton-Proton Collisions at 
$\sqrt{s} = 13$ TeV. 2023. Available online: \url{https://doi.org/10.17182/hepdata.141028} (accessed on 13 July 2023).

\bibitem{NIMA} Ranucci, G. The profile likelihood ratio and the look elsewhere effect in high energy physics.  {\em Nucl. Inst. Meth. Phys. Res. A} {\bf 2012}, {\em 661}, 77-85.  

\bibitem{three} Particle Data Group. Review of Particle Physics.  {\em Phys. Lett. B} {\bf 2008}, {\em 667}, 1.



\end{thebibliography}
\end{document}